\documentclass[aps,pre,reprint,superscriptaddress,longbibliography]{revtex4-1}
\usepackage{scrextend} 
\usepackage{amsmath,amssymb,graphicx,subfigure,color,times,tabularx,hyperref,fancyhdr}
\usepackage{esint}
\begin{document}
\title{The Madelung Problem of Finite Crystals}
\author{Yihao Zhao}
\affiliation{Key Laboratory of Laser \& Infrared System of Ministry of Education, Shandong University, Qingdao 266237, P. R. China}
\affiliation{Qingdao Institute for Theoretical and Computational Sciences (QiTCS), Center for Optics Research and Engineering, Shandong University, Qingdao 266237, P. R. China}
\author{Yang He}
\affiliation{Key Laboratory of Laser \& Infrared System of Ministry of Education, Shandong University, Qingdao 266237, P. R. China}
\affiliation{Qingdao Institute for Theoretical and Computational Sciences (QiTCS), Center for Optics Research and Engineering, Shandong University, Qingdao 266237, P. R. China}
\author{Zhonghan Hu} \email{zhonghanhu@sdu.edu.cn}
\affiliation{Key Laboratory of Laser \& Infrared System of Ministry of Education, Shandong University, Qingdao 266237, P. R. China}
\affiliation{Qingdao Institute for Theoretical and Computational Sciences (QiTCS), Center for Optics Research and Engineering, Shandong University, Qingdao 266237, P. R. China}
\begin{abstract}
The Coulomb potential at an interior ion in a finite crystal of size $p$ is given by a linear superposition of contributions from displacement vectors ${\mathbf r}=(x,y,z)$ to its neighbors. 
This additive structure underlies universal relationships among Madelung constants and applies to both standard periodic boundary conditions and alternative Clifford supercells.
Each pairwise contribution decomposes into three physically distinct components: a periodic bulk term, a quadratic boundary term, and a finite-size correction whose leading order term is $[24r^4-40(x^4+y^4+z^4)]/[9\sqrt{3} (2p+1)^2]$ for cubic crystals with unit lattice constant.
Combining this decomposition with linear superposition yields a rapidly convergent direct-summation scheme, accurate even at $p=1$ ($3^3$ unit cells), enabling hands-on calculations of Madelung constants for a wide range of ionic crystals. 
\end{abstract} \maketitle

\section{Introduction}
The classical Madelung problem concerns the electrostatic potential at the site of a reference ion in an infinite ionic crystal (e.g., NaCl or CsCl)\cite{Madelung1918}. When scaled by the potential of an isolated nearest-neighbor counterion, 
this defines the Madelung constant---a dimensionless quantity central to the lattice energy per ion pair. 
Its computation, however, is mathematically subtle: the conditionally convergent nature of the Coulomb series renders naive direct summation (e.g., over spherical or cubic shells) not only slow but also ambiguous, as the result may depend on the summation order
(e.g.\cite{DeLeeuw_Smith1980,Smith1981,Smith2008,Ballenegger2014,Hu2014ib,Pan_Hu2017,Zhao_Hu2025}).

A rich literature has emerged to address this challenge.
At one extreme lie highly accurate but analytically involved integral-transform methods---most notably Ewald summation and its related approaches\cite{Ewald1921,Nijboer1957,DeLeeuw_Smith1980,Borwein1985}.
At the other are conceptually simpler direct-summation strategies: charge-renormalization schemes (e.g., Evjen's surface-weighting\cite{Evjen1932}, Harrison's neutralizing shell\cite{Harrison2006}, or multipole-cancellation approaches\cite{Marathe1983,Sousa1993,Derenzo2000,Gelle2008}) 
and geometric alternatives like the Clifford supercell (CS) method\cite{Tavernier2020,Tavernier2021}, which redefines distances using the intrinsic metric of a Clifford torus.

\begin{table}[!htb]   
\caption{Madelung constants for CsCl (${\mathbf v}_1=(0.5, 0.5, 0.5)$) computed using the explicitly corrected direct sum [Eqs.~\eqref{eq:pw} to~\eqref{eq:corr}] and the CS method [Eqs.~\eqref{eq:rd} and \eqref{eq:cspw}], for varying supercell sizes.
The reference value is 1.76\,267\,477\,307\,098\cite{csclseq}.}
\label{tab:cscl}
\begin{tabular}{ccl|ccl}\hline $ p$ & $\sqrt{3}\nu_{\rm pbc}({\mathbf v}_1)/2$ &    $\,$ abs. error $\,$& $K$     &   $\sqrt{3}\nu_{\rm CS}({\mathbf v}_1)/2$ & $\,$ abs. error \\[0.4ex] \hline
                                    1           &               1.7629780255                  & \,\,3.0 $\times 10^{-4}$          &   3      &    1.408794                    & -3.5$\times 10^{-1}$ \\[0.5ex]
                                    5           &               1.7626721815                  &    -2.6 $\times 10^{-6}$          &  11      &    1.744082                    & -1.9$\times 10^{-2}$\\[0.5ex]
                                   20           &               1.7626747599                  &    -1.3 $\times 10^{-8}$          &  41      &    1.761379                    & -1.3$\times 10^{-3}$\\[0.5ex]
                                   60           &               1.7626747729                  &    -1.7 $\times 10^{-10}$         & 121      &    1.762526                    & -1.5$\times 10^{-4}$\\\hline
\end{tabular}\end{table}
Yet truly analytical, hands-on computation remains elusive.
For CsCl, Evjen's method---reducing surface charges by $1/2$, $1/4$, and $1/8$ on faces, edges, and corners---fails to converge to the correct value; 
Harrison's neutralization---enforcing neutrality via a surrounding charged spherical shell---converges correctly but slowly, requiring radii of hundreds of lattice constants (i.e., millions of unit cells) to reach $\sim 10^{-3}$ accuracy. 
While the multipole-cancellation scheme by Gelle and Lepetit achieves exponential convergence\cite{Gelle2008}, it relies on increasingly intricate charge distributions that complicate practical implementation.
The CS method, though conceptually simple and exhibiting ${\cal O}(K^{-2})$ convergence with a $K\times K\times K$ supercell, still demands $K\geqslant 40$ (tens of thousands of unit cells) for modest ($\sim 10^{-3}$) accuracy, as Tab.~\ref{tab:cscl} illustrates.

\begin{figure}[!htb]\centerline{\includegraphics[width=7.7cm]{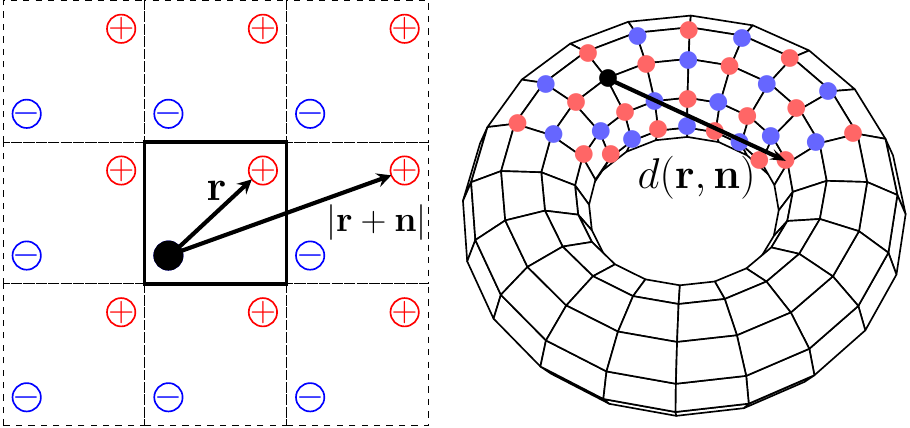} }          
\caption{Two-dimensional cross-sectional view of two particles with unit charges ($\pm 1$) separated by vector ${\mathbf r}=(x,y,z)$ and their periodic images under standard periodic boundary conditions (left) and in the CS method (right).
For the conventional direct sum, the potential at the reference particle (black dot) is obtained by summing the Coulomb interactions $1/\left| {\mathbf r}+{\mathbf n} \right|$ from all periodic images, as described by Eq.~\eqref{eq:pw}.
In the CS method ($K^3$ unit cells with unit side length), the conventional Euclidean distance $\left| {\mathbf r} + {\mathbf n}\right|$ is replaced by the renormalized distance $d(\mathbf{r}, \mathbf{n})$ (Eqs.~\eqref{eq:rd} and~\eqref{eq:cspw}).\label{fig:rd}}\end{figure}
In this work, we resolve the long-standing ambiguity in the Madelung problem or the Coulomb lattice sum---stemming from the conditional convergence of the electrostatic lattice sum---by rigorously defining and analytically separating boundary and finite-size effects in finite crystals.
Our approach achieves rapid convergence from conventional direct summation by incorporating higher-order multipolar effects via exact boundary and finite-size corrections---without resorting to charge or distance renormalization.
For the CsCl-like two-particle configuration depicted in Fig.~\ref{fig:rd}, the direct Coulomb sum for a finite crystal comprises of a shape- and size-independent bulk term $\nu_{\rm pbc}({\mathbf r})$, 
a size-independent boundary contribution $\nu_{\rm b}({\mathbf r}|{\mathbf s})$, and a finite-size correction $\nu_{\rm corr}({\mathbf r},p|{\mathbf s})$
\begin{equation} \begin{aligned} \nu({\mathbf r},p|{\mathbf s})& = \frac{1}{r} + \sum_{ {\mathbf n}\neq {\mathbf 0} }^{{\cal L}(p|{\mathbf s})} \left[ \frac{1}{\left| {\mathbf r} + {\mathbf n} \right | }  - \frac{1}{n} \right] \\
                                         & = \nu_{\rm pbc}({\mathbf r}) + \nu_{\rm b}({\mathbf r}|{\mathbf s}) + \nu_{\rm corr}({\mathbf r},p|{\mathbf s})  \end{aligned} \label{eq:pw}, \end{equation}
where ${\mathbf r}=(x,y,z)$ denotes the relative displacement, ${\mathbf n}=(n_x,n_y,n_z)$ is the lattice vector, $n=|{\mathbf n}|$, and ${\cal L}(p|{\mathbf s})$ the set of all lattice vectors in a finite crystal of linear extent $p$ (with $K=2p+1$ unit cells per dimension) and shape ${\mathbf s}$. 
Here ${\mathbf s}=(0, 0, 0)$ indicates the shape of the finite cubic crystal.
The meaning of ${\mathbf s}$ in a general orthogonal lattice will be clarified later.
In the present explicit finite-size correction (EC) method, we isolate $\nu_{\rm pbc}$, which is the bulk contribution under the periodic boundary condition (PBC),
from the direct sum $\nu({\mathbf r},p|{\mathbf s})$ and then scale it by the potential of an isolated counterion at the nearest-neighbor distance to obtain the Madelung constant.
The EC method thus provides a well-defined approximation to the size- and shape-independent bulk term $\nu_{\rm pbc}$ by evaluating a finite Coulomb lattice sum supplemented with explicit boundary and finite-size corrections.
Eq.~\eqref{eq:pw} captures the essential physical meaning of the Madelung constant as originally conceived by Ewald\cite{Ewald1921}, namely the electrostatic potential at a reference ion due to all other ions in the crystal---distinct from the response of the crystal to an externally applied field\cite{Rayleigh1892}.
It has long been recognized that conditionally convergent lattice sums can yield different results depending on the summation order or the geometric perspective adopted\cite{DeLeeuw_Smith1980}.
The key advance of the present work lies in its unambiguous definition of both size and shape for a finite crystal.
In contrast, previous approaches---such as those based on infinite lattice sums (e.g., Ref.\cite{Hu2014ib} and references therein)---leave the boundary condition ill-defined, thereby obscuring the physical origin of shape-dependent contributions.

For a cubic crystal in a simple cubic lattice (lattice constants $l_x=l_y=l_z=1$), the boundary and finite-size corrections admit closed-form expressions:
\begin{equation} \nu_{\rm b}({\mathbf r}|{\mathbf s}) = -2\pi r^2/3,  \label{eq:bound} \end{equation}
and
\begin{equation} \nu_{\rm corr}({\mathbf r},p|{\mathbf s}) = \frac{24r^4 - 40(x^4+y^4+z^4)}{9\sqrt{3} (2p+1)^2} + {\cal O}(p^{-4}) \label{eq:corr}. \end{equation}
In contrast, the CS method employs the renormalized distance\cite{Aragao2019,Tavernier2020}
\begin{equation}  d({\mathbf r},{\mathbf n}) = \frac{K}{\sqrt{2}\pi}\left[ 3 - \sum_{\alpha=1}^3 \cos \frac{2\pi ({\mathbf r}+{\mathbf n})\cdot {\mathbf e}_\alpha}{K} \right]^{1/2} \label{eq:rd}, \end{equation}
with ${\mathbf e}_\alpha$ the Cartesian unit vectors, and evaluates the bulk potential via the uncorrected direct sum
\begin{equation} \nu_{\rm CS}({\mathbf r}) = \frac{1}{d({\mathbf r},{\mathbf 0})} + \sum_{{\mathbf n}\neq {\mathbf 0}}^{{\cal L}(p|{\mathbf 0})}\left[ \frac{1}{d({\mathbf r},{\mathbf n})} - \frac{1}{d({\mathbf 0},{\mathbf n})} \right] \label{eq:cspw}. \end{equation}
As demonstrated in Tab.~\ref{tab:cscl} for CsCl, incorporating Eqs.~\eqref{eq:bound} and~\eqref{eq:corr} accelerates convergence from ${\cal O}(K^{-2})$ (CS) to ${\cal O}(K^{-4})$ (EC).
Remarkably, the EC method achieves $3\times 10^{-4}$ accuracy with the minimal supercell ($p=1, K=3$), and delivers nine-digit precision at $p=60$---all without renormalization.

\begin{table}[!htb]   
\caption{Bulk contributions for displacement vectors ${\mathbf v}_2 = (0.5, 0, 0)$, ${\mathbf v}_3 = (0.5, 0.5, 0)$, and ${\mathbf v}_4 = (0.25, 0.25, 0.25)$ computed using the EC method and referenced to $p=60$ result (nine significant digits).}\label{tab:vec3}
\begin{tabular}{cccc}\hline $p$ & \quad\quad$\nu_{\rm pbc}({\mathbf v}_2)$\quad\quad &\quad\quad $\nu_{\rm pbc}({\mathbf v}_3)$\quad\quad & \quad\quad$\nu_{\rm pbc}({\mathbf v}_4)$\quad\quad \\[0.4ex] \hline
                            1   &              2.741146342                           &        2.255022524                                 &  2.636876604                       \\[0.5ex]
                            5   &              2.741365130                           &        2.254776731                                 &  2.636813436                       \\[0.5ex]
                           20   &              2.741365174                           &        2.254775952                                 &  2.636813487                       \\[0.5ex]
                           ref. &              2.741365175                           &        2.254775948                                 &  2.636813487                       \\\hline
\end{tabular} \end{table}
Table~\ref{tab:vec3} shows that, the correction scheme maintains high accuracy (better than $10^{-8}$ at $p=20$) across a range of displacement vectors, confirming its robustness for arbitrary displacements.
As detailed later, the two-ion configuration serves as a fundamental building block for the Madelung problem: for crystals with multi-atom unit cells (e.g., the crystal structures in Tab.~\ref{tab:unit}), 
the total Madelung constant is obtained by linearly superposing contributions from all symmetry-inequivalent ion pairs, each evaluated via the direct sum in Eq.~\eqref{eq:pw} and corrected analytically using $\nu_{\rm b} + \nu_{\rm corr}$.
This additive property, combined with the ${\cal O}(p^{-4})$ convergence of every term, enables a truly analytical, renormalization-free evaluation of Madelung constants---even for complex ionic structures such as NaCl, ZnS, CaF$_2$ and CaTiO$_3$.

\begin{table}[!htb]   
\caption{Reduced coordinates of ions in a cubic unit cell of a typical rocksalt (Na$_4$Cl$_4$), zincblende (Zn$_4$S$_4$) and fluorite (Ca$_4$F$_8$), with the unit cell scaled to unit length: $ l_x = l_y = l_z = 1 $.}\label{tab:unit}
\begin{tabular}{cccc}\hline  $4$Na$^+$/$4$Ca$^{2+}$           &        $4$Cl$^-$/$4$Zn$^{2+}$          &            $4$S$^{2-}$/$4$F$^-$                    &          $4$F$^-$        \\[0.5ex]
                                  (0, 0, 0)     &  (0.5, 0, 0)          &    (0.25, 0.25, 0.25) &    (0.25, 0.25, 0.75)\\[0.5ex]
                                   (0.5, 0, 0.5) & (0, 0.5, 0)        &(0.75, 0.25, 0.75)      & (0.25, 0.75, 0.25)  \\[0.5ex]
                                   (0.5, 0.5, 0) & (0, 0, 0.5)        & (0.75, 0.75, 0.25)      & (0.75, 0.25, 0.25) \\ [0.5ex]
                                  (0, 0.5, 0.5)  & (0.5, 0.5, 0.5)& (0.25, 0.75, 0.75)& (0.75, 0.75, 0.75)\\\hline
\end{tabular} \end{table}
In the following, we first illustrate the essential meaning of Eq.~\eqref{eq:pw} for a simple cubic lattice by comparing two cubic crystals in Section~\ref{sec:twocube} and by comparing cubic and cuboid crystals in Section~\ref{sec:cuboid}, respectively. 
Rigorous derivations of $\nu_{\rm b}({\mathbf r}|{\mathbf s})$ and $\nu_{\rm corr}({\mathbf r},p|{\mathbf s})$ for a general orthogonal lattice are provided in Section~\ref{sec:general}.
While $\nu_{\rm b}({\mathbf r}|{\mathbf s})$ admits a closed-form expression for crystals with arbitrary aspect ratios, the explicit closed-form expression $\nu_{\rm corr}({\mathbf r},p|{\mathbf s})$ in Eq.~\eqref{eq:corr} is derived for cubic crystals in a cubic lattice.
Nevertheless, Section~\ref{sec:app} demonstrates that this formulation is sufficient to compute highly accurate Madelung constants for a broad set of benchmark structures---requiring only minimal computational effort. 
Finally, we draw conclusions in Section~\ref{sec:con} and provide evaluations of selected integrals in the Appendix and in the Supporting Information.
\section{Difference between two cubic crystals}\label{sec:twocube}
\begin{figure}[!htb]\centerline{\includegraphics[width=5cm]{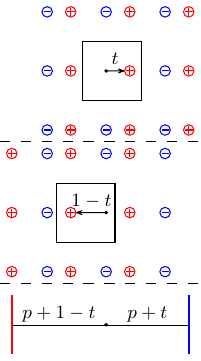} }          
\caption{
Cross-sectional view of two cubic crystals of size $p$, each composed of unit charges ($\pm 1$), separated by the vectors $(t, 0, 0)$ (top) and $(t-1, 0, 0)$ (middle), respectively. In the limit of large $p$, the difference of charges in their outermost layers can be approximated by a pair of uniformly charged parallel plates
located at the planes $x=-(p+1-t)$ and $x=p+t$ (bottom).
\label{fig:twocube}}\end{figure}
Eq.~\eqref{eq:pw} decomposes the lattice sum over long-range Coulomb interactions into three distinct components: (i) the bulk term $\nu_{\rm pbc}({\mathbf r})$ which is a periodic function of ${\mathbf r}$, (ii) the boundary term $\nu_{\rm b}({\mathbf r}|{\mathbf s})$, which depends on the shape of the crystal and is not a
periodic function of ${\mathbf r}$, (iii) the
finite-size correction $\nu_{\rm corr}({\mathbf r},p|{\mathbf s})$, which vanishes at the limit $p\to\infty$.
Obviously, if the interactions were short-ranged---decaying sufficiently fast or vanishing beyond a certain distance---the boundary term would not appear.
In such cases, the bulk term accounts for the influence of nearby particles, while the finite-size correction captures the effect of the finite extent of the crystal.
The shape-dependent boundary term has attracted considerable attention in the literature (see, e.g.\cite{DeLeeuw_Smith1980,Smith1981,Smith2008,Ballenegger2014,Hu2014ib}). 
Previous derivations often relied on intricate integral transforms. In contrast, we present here a transparent derivation of the boundary effect using only elementary calculus and basic electrostatics, based on a direct comparison of two finite crystals. 

To this end, we first consider two cubic crystals in a simple cubic lattice with unit lattice constant. 
Their relative displacements are taken as ${\mathbf r}_1 = (t, 0, 0)$ and its periodic translation: ${\mathbf r}_2 = (t-1, 0, 0)$, respectively.
Both crystals have identical dimensions $(2p+1)\times(2p+1)\times(2p+1)$, differing only in the outermost layers perpendicular to the $x$ direction, as illustrated in Fig.~\ref{fig:twocube}.
For large $p$, the finite-size correction becomes negligible, and the local environment around the central ion (marked by dot in Fig.~\ref{fig:twocube}) is identical in both crystals except at the boundaries.
When the central ion is placed at the origin, the difference in the electrostatic potential it experiences arises solely from the two layers located at $x=p+t$ and $x=-(p+1-t)$.
Because $p$ is large, these layers can be approximated as uniformly charged square plates of side length $a=2p+1$ and charge densities $\sigma = \pm 1$.

The electrostatic potential produced by a single square plate of size $a$ and charge density $\sigma$, evaluated at a point on its central axis at distance $z$, is given by $\sigma\,\phi(z,a)$ \cite{Hummer1996,Ciftja2020}, where
\begin{equation} \phi(z,a) = 2 a\, \psi(2z/a)  ,\end{equation}
and the dimensionless function $\psi(x)$ is even in $x$ and defined as
\begin{equation} \psi(x) = \log \frac{\sqrt{x^2+2}+1}{\sqrt{x^2+2} - 1} - x\, {\rm atan}\frac{1}{x\sqrt{x^2+2}} , \label{eq:psi} \end{equation}
with ${\rm atan}(\cdot)$ denoting the inverse tangent function.
With $z_1 = p+1-t$ and $z_2 = p+t$ for $t\in (0,1)$, the boundary contribution can be expressed as
\begin{equation} \nu_{\rm b}({\mathbf r}_2|{\mathbf s}) - \nu_{\rm b}({\mathbf r}_1|{\mathbf s}) = \lim_{p\to \infty} \phi(z_1,a) - \phi(z_2,a)\end{equation}

To verify this relation, note that
\begin{equation} \frac{2z_1}{a} = 1 + \frac{1-2t}{a}; \quad \frac{2z_2}{a} = 1 - \frac{1-2t}{a} .\end{equation}
Expanding $\psi(\tau+x)$ around $x=0$ to second order yields
\begin{multline} \psi(\tau+x) = \psi(\tau) - x\, {\rm atan}\frac{1}{\tau\sqrt{\tau^2+2}} \\ + \frac{x^2}{\left(\tau^2+1\right)\sqrt{\tau^2+2}} + \cdots . \label{eq:taylor}\end{multline}
Substituting $\tau = 1$ and $x = \pm (1-2t)/a$, we obtain
\begin{equation} \lim_{p\to \infty} \phi(z_1,a) - \phi(z_2,a)  = - 4 a \frac{\pi}{6}\frac{1-2t}{a}  = \frac{2\pi}{3}\left(2t-1\right) .\end{equation}
This result matches the difference obtained directly from the boundary term in Eq.~\eqref{eq:bound}:
\begin{equation} \nu_{\rm b}({\mathbf r}_2|{\mathbf s}) - \nu_{\rm b}({\mathbf r}_1|{\mathbf s})  = -\frac{2\pi}{3}\left[ (t-1)^2 - t^2 \right] = \frac{2\pi}{3}\left(2t-1\right) . \end{equation}
Thus, the boundary effect derived from the electrostatics of parallel plates agrees exactly with the analytical expression, confirming the validity of the approach.
\section{Difference between cubic and cuboid crystals}\label{sec:cuboid}
\begin{figure}[!htb]\centerline{\includegraphics[width=7.8cm]{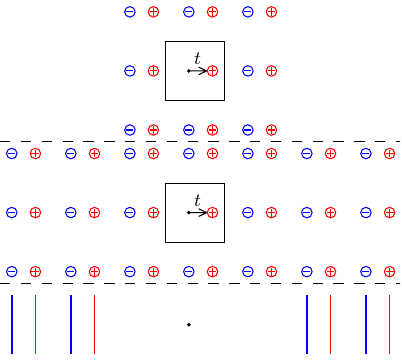} }          
\caption{
Cross-sectional view of cubic (top) and cuboid (middle) crystals of size $p$, both composed of unit charges ($\pm 1$), separated by the vectors $(t, 0, 0)$. For large $p$, the additional layers presented in the cuboid crystal can be approximated by pairs of uniformly charged parallel plates
located at the planes $x=-(3p+1)$, $-(3p+1-t)$, $\cdots$, $-(p+1)$, $-(p+1-t)$, $p+1$, $p+1+t$, $\cdots$, $3p+1$, $3p+1+t$ (bottom).
\label{fig:cuboid}}\end{figure}
By comparing two cubic crystals, the preceding section provided a tutorial demonstration that the boundary term $\nu_{\rm b}({\mathbf r}|{\mathbf s})$ is not a periodic function of ${\mathbf r}$. In this section, we extend the analysis to compare cubic and cuboid crystals in a simple cubic lattice with unit lattice constant. This comparison illustrates that the boundary term also depends on the shape of the crystal.

In both crystals, the two charges are displaced by the same vector ${\mathbf r}_1 = (t,0,0)$. 
The cubic crystal of size $p$ is identical to the one considered previously.
The cuboid crystal, by contrast, has dimensions $3(2p+1)\times (2p+1) \times (2p+1)$, i.e., it is three times longer in the $x$-direction.
Fig.~\ref{fig:cuboid} shows a cross-sectional view of the two crystals in the $xy$-plane, along with their difference approximated as pairs of parallel plates.

For large $p$, each of the $(2p+1)$ additional layers in the cuboid crystal---indexed by $k=p+1, p+2, \cdots, 3p+1$---can be represented by two pairs of oppositely charged parallel plates.
Each plate is perpendicular to the $x$-axis, has side length $a=2p+1$, and carries a charge density $\sigma=\pm 1$. 
The contribution from the $k$-th layer consists of four terms:
$2a\psi_+(k)$, $2a\psi_-(k)$, $-2a \psi_0(k)$, $-2a \psi_0(k)$ where
\begin{equation} \psi_+(k) = \psi\left( \frac{2k+2t}{a} \right), \end{equation}
\begin{equation} \psi_-(k) = \psi\left( \frac{2k-2t}{a} \right), \end{equation}
\begin{equation} \psi_0(k) = \psi\left( \frac{2k}{a} \right). \end{equation}
and $\psi(x)$ is defined in Eq.~\eqref{eq:psi}.

Let ${\mathbf s}_1$ and ${\mathbf s}_2$ denote the shapes of  the cubic and cuboic crystals, respectively. 
The boundary effect is fully captured by the following relation:
\begin{multline} \nu_{\rm b}({\mathbf r}_1|{\mathbf s}_2) - \nu_{\rm b}({\mathbf r}_1|{\mathbf s}_1) \\  = \lim_{p\to\infty} \sum_{k=p+1}^{3p+1} 2a \left[\psi_+(k) + \psi_-(k) - 2 \psi_0(k) \right]. \label{eq:dcuboid} \end{multline}
Substituting $\tau = 2k/a$ and $x = \pm 2t/a$ into the Taylor series in Eq.~\eqref{eq:taylor}, the right hand side (rhs) of Eq.~\eqref{eq:dcuboid} becomes
\begin{equation} \text{rhs.} = \lim_{p\to\infty} \sum_{k=p+1}^{3p+1} \frac{16 at^2}{ \left(4k^2 + a^2 \right) \sqrt{ 4k^2 + 2a^2} }. \end{equation}
Replacing the sum by an integral with the substitution $x=2k/a$ yields,
\begin{equation} \text{rhs.}= 8t^2 \int_1^3 \frac{dx}{{\left(x^2+1\right)\sqrt{x^2+2} }}. \end{equation}
As shown in the Appendix, the indefinite integral evaluates to 
\begin{equation} \int \frac{dx}{{\left(x^2+1\right)\sqrt{x^2+2} }} ={\rm atan}\frac{x}{\sqrt{x^2+2}} + \text{const.}, \end{equation}
so that
\begin{equation}  \nu_{\rm b}({\mathbf r}_1|{\mathbf s}_2) - \nu_{\rm b}({\mathbf r}_1|{\mathbf s}_1) = 8t^2 \left[ {\rm atan}\frac{3}{\sqrt{11}} - {\rm atan}\frac{1}{\sqrt{3}} \right]. \label{eq:dcuboid2} \end{equation}
This result explicitly confirms that the boundary term depends on the crystal shape. The validity of the above equation will be further verified in the next section, where we derive the explicit expression for the boundary term of a cuboid crystal.

\section{General results for an orthogonal lattice}\label{sec:general}
It is well known that the Coulomb lattice sum is a conditionally convergent series whose value depends on the macroscopic shape of the crystal\cite{DeLeeuw_Smith1980,Smith1981}.
A rigorous treatment therefore requires defining a sequence of finite crystals that grow in size while maintaining a fixed geometric shape.
We consider an orthogonal lattice where the lattice constants $l_x$, $l_y$, and $l_z$ represent the edge lengths of the orthorhombic unit cell.
For a centro-symmetric crystal composed of one central primitive cell and $ N_1 $, $ N_2 $, $ N_3 $ replicated cells extending symmetrically along the $\pm x$, $\pm y$, and $\pm z$-directions, respectively, the position of each unit cell is specified by the lattice vector
\begin{equation} \mathbf{n} = (n_x, n_y, n_z) = (n_1 l_x, n_2 l_y, n_3 l_z), \end{equation}  
where $ n_1, n_2, n_3 $ are integers in the range $-N_\alpha \leqslant n_\alpha \leqslant N_\alpha$ for $\alpha=1,2,3$.
The overall dimensions of the resulting cuboid are therefore $a=(2N_1+1)l_x$, $b=(2N_2+1)l_y$, and $c=(2N_3+1)l_z$.

In the isotropic case where $ N_1 = N_2 = N_3 = p $, the crystal retain the same aspect ratios as the primitive cell:
\begin{equation} a : b : c = l_x : l_y : l_z. \end{equation}
Thus, as $p$ increases from zero,  the sequence of crystals maintains a fixed shape under isotropic growth.

For the general anisotropic case, we define an effective size $p$ such that $2p+1$ equals the greatest common divisor of $2N_1+1$, $2N_2+1$ and $2N_3+1$. 
This condition implies the simultaneous factorizations:
\begin{equation} 2N_\alpha + 1 = (2s_\alpha + 1)(2p+1) \quad\text{for}\quad \alpha = 1,2,3, \end{equation}
where the quotients ($2s_1+1$, $2s_2+1$, $2s_3+1$) form a co-prime triplet---i.e., their only common divisor is $1$.
As $p$ increases with fixed ${\mathbf s}=(s_1,s_2,s_3)$, the crystal dimensions evolve as:
$a=(2p+1)(2s_1+1)l_x$, $b=(2p+1)(2s_2+1)l_y$, and $c=(2p+1)(2s_3+1)l_z$, giving invariant aspect ratios:
\begin{equation} a:b:c =(2s_1+1)l_x \,:\, (2s_2+1)l_y \,:\, (2s_3+1)l_z .\end{equation}
Consequently, all crystals in this sequence maintain an identical geometric shape, uniquely specified by the parameter ${\mathbf s}$ for any given set of lattice constants $(l_x,l_y,l_z)$.

Now, suppose each unit cell consists of $M$ ions with charges $q_j$ located at positions ${\mathbf r}_j$ for $j=1,2,\cdots,M$. 
The electric potential at the $i$-th ion in the primary cell (${\mathbf n}={\mathbf 0}$) is then given by
\begin{equation} \phi_p(i) = \sum_{j=1,j\neq i}^M  \frac{q_j}{ \left| {\mathbf r}_j - {\mathbf r}_i \right| } + \sum_{j = 1}^M \sum_{ {\mathbf n}\neq{\mathbf 0} }^{{\cal L}(p|{\mathbf s})} \frac{q_j}{ \left| {\mathbf r}_j  + {\mathbf n} - {\mathbf r}_i \right| }. \end{equation}
This expression---and analogous sums for the electrostatic energy---constitutes the starting point for deriving the Ewald summation and the associated shape-dependent boundary contributions in the limit of a macroscopic (infinite) crystal (e.g.\cite{DeLeeuw_Smith1980,Smith1981,Ballenegger2014,Hu2014ib}).
In contrast, we rigorously define the geometry of finite crystals for a general orthogonal lattice, thereby enabling a transparent and systematic treatment of finite-size effects.
In the absence of such a formulation, the boundary contribution in the infinite limit remains ambiguous---often obscured by implicit assumptions and lacking direct justification from finite, physically realizable systems.

Under the condition of charge neutrality, $\sum_{j=1}^M q_j = 0$, the self-interaction part ($i=j$) in the second sum can be rewritten as
\begin{equation} \sum_{ {\mathbf n}\neq{\mathbf 0} }^{{\cal L}(p|{\mathbf s})} \frac{q_i}{\left| {\mathbf n} \right|} = - \sum_{j=1,j\neq i}^M \sum_{ {\mathbf n}\neq{\mathbf 0} }^{{\cal L}(p|{\mathbf s})} \frac{q_j}{\left| {\mathbf n} \right|}, \end{equation}
allowing the potential to be recast in a pairwise form\cite{Hu2014ib,Yi_Hu2017pairwise,Zhao_Hu2025}
\begin{equation} \phi_p(i) = \sum_{j=1,j\neq i}^M  q_j \nu({\mathbf r}_j-{\mathbf r}_i,p|{\mathbf s}) \label{eq:phi}, \end{equation}
where the effective pairwise interaction $\nu({\mathbf r},p|{\mathbf s})$ is defined in Eq.~\eqref{eq:pw}.

In prior work, the infinite-system limit of this interaction is decomposed as\cite{Hu2014ib,Yi_Hu2017pairwise,Zhao_Hu2025}
\begin{equation} \nu({\mathbf r},\infty|{\mathbf s}) = \nu_{\rm pbc}({\mathbf r}) + \nu_{\rm b}({\mathbf r}|{\mathbf s}), \end{equation}
where $\nu_{\rm pbc}({\mathbf r})$ is geometry-independent and represents the bulk contribution.
This term has served as a foundation for analytical studies of structural correlations in bulk electrolytes and dielectrics\cite{Hu2022,Gao_Hu2023}, as well as dielectric responses at interfaces\cite{Hu2014spmf,Pan_Hu2017,Pan_Hu2019}.
It has been demonstrated recently that\cite{Zhao_Hu2025}, the pairwise sum over
 $ \nu_{\rm pbc}({\mathbf r})$ provides critical insight into the role of the neutralizing background charge and enables a consistent formulation of energies and pressure in systems comprising both discrete and continuous charge distributions\cite{Onegin2024,Li2025,Demyanov2025}.
The second term, $\nu_{\rm b}({\mathbf r}|{\mathbf s})$  accounts for the arrangement of charges at the infinite boundary\cite{Zhao_Hu2025}. 
Crucially, as we demonstrate below, it remains a non-negligible contribution even for finite crystals.
In reciprocal space, it can be expressed as\cite{Hu2014ib,Zhao_Hu2025} 
\begin{equation} \nu_{\rm b}({\mathbf r}|{\mathbf s}) = -\frac{2\pi}{V} \lim_{ {\mathbf k}| {\mathbf s} \to 0 } \frac{\left( {\mathbf k}\cdot {\mathbf r} \right)^2}{\left| {\mathbf k} \right| ^2}, \end{equation}
where $V=l_xl_yl_z$ is the volume of a unit cell and the limit is taken such that the wavevector ${\mathbf k}$ approaches zero along a direction determined by the aspect ratios encoded in ${\mathbf s}$.
This representation makes manifest that $\nu_{\rm b}({\mathbf r}|{\mathbf s})$ is always non-positive. 
Equivalently, in real space, $\nu_{\rm b}({\mathbf r}|{\mathbf s})$ admits an electrostatic interpretation: it is the interaction energy between a point dipole at the origin and a uniformly polarized continuum filling the crystal volume\cite{Smith2008,Ballenegger2014,Pan2017,Zhao_Hu2025}:
\begin{equation} \nu _{\rm b}({\mathbf r}|{\mathbf s}) = \frac{1}{2V} \int_{\Omega(p|{\mathbf s})} d{\mathbf x} \, \left({\mathbf r}\cdot \nabla_{\mathbf x}\right)^2\frac{1}{\left| {\mathbf x}\right|}, \label{eq:ib} \end{equation} 
where the integration domain $\Omega(p|{\mathbf s})$ denotes the total volume of the crystal:
\begin{equation} \Omega(p|{\mathbf s}) = abc = V (2p+1)^3 \Pi_{\alpha = 1}^3 (2s_\alpha+1). \end{equation}
Notably, although this domain depends on $p$, explicit evaluation [see Supporting Information (SI)] yields a closed-form, $p$-independent expression:
\begin{equation} \nu _{\rm b}({\mathbf r}|{\mathbf s}) = - \frac{4}{V}\sum_{\alpha=1}^3 \left({\mathbf r}\cdot{\mathbf e}_\alpha\right)^2 {\rm atan}\frac{1}{\xi_\alpha^2\sqrt{\xi_1^2+\xi_2^2+\xi_3^2}} \label{eq:ibabc}, \end{equation}
where the dimensionless parameter $\xi_\alpha$ are defined by normalizing each side length with the geometric mean of the three:
\begin{equation}\xi_1 = \frac{a}{(abc)^{1/3}}; \quad \xi_2 = \frac{b}{(abc)^{1/3}};\quad \xi_3 = \frac{c}{(abc)^{1/3}}. \end{equation}
For a cubic crystal with unit lattice constant ($l_x=l_y=l_z=1$), $\nu _{\rm b}({\mathbf r}|{\mathbf s})$ of Eq.~\eqref{eq:ibabc} simplifies to Eq.~\eqref{eq:bound}. 
Inserting Eq.~\eqref{eq:ibabc} into the left hand side of Eq.~\eqref{eq:dcuboid2} confirms the validity of Eq.~\eqref{eq:dcuboid2} (see also Eq.~\eqref{eq:ibapp} in the Appendix).

For a finite crystal of size $p$, the effective pairwise interaction $\nu({\mathbf r},p|{\mathbf s})$ deviates from $\nu({\mathbf r},\infty|{\mathbf s})$ by the finite-size correction $\nu_{\rm corr}({\mathbf r},p|{\mathbf s})$ introduced in Eq.~\eqref{eq:pw}.
$\nu_{\rm corr}({\mathbf r},p|{\mathbf s})$ accounts for the omitted contributions from lattice vectors outside the finite summation domain. Explicitly,
\begin{equation} \nu_{\rm corr}({\mathbf r},p|{\mathbf s}) =  - \sum_{ {\mathbf n}\notin {\cal L}(p|{\mathbf s}) }^{{\cal L}(\infty|{\mathbf s})} \left[ \frac{1}{\left| {\mathbf r} + {\mathbf n} \right | }  - \frac{1}{n} \right],  \label{eq:corrdef} \end{equation}
with the sum extending over all lattice vectors excluded from the finite crystal but present in the infinite lattice of the same shape.
 
To estimate $\nu_{\rm corr}({\mathbf r},p|{\mathbf s})$ to a desired accuracy, we expand the summand for distant cells ($r \ll n $) as a multipolar series:
\begin{equation} \frac{1}{\left| {\mathbf r} + {\mathbf n} \right | }  - \frac{1}{n} =\sum_{k=1}^\infty \frac{ (-r)^k P_k(\cos\theta) }{n^{k+1}} =  \sum_{k=1}^\infty \frac{\left({\mathbf r}\cdot \nabla_{\mathbf n} \right)^k}{k!} \frac{1}{n}, \label{eq:mpole} \end{equation}
where $\cos\theta = {\mathbf r}\cdot {\mathbf n}/(nr)$, $P_k(\cos\theta)$ is the $k$-th Legendre polynomial, and $\nabla_{\mathbf n}$ is the gradient operator with respect to the lattice vector ${\mathbf n}$.
While multipolar series underpins fast algorithms for evaluating potentials in Coulomb systems\cite{Greengard1987,Greengard1988,Kudin1998}, we instead provide an analytical treatment of the boundary term and asymptotic finite-size behavior for finite crystals with explicitly specified shape and size.
Gelle and Lepetit performed a similar multipole expansion using spherical coordinates rather than Cartesian coordinates\cite{Gelle2008}. 
Their approach facilitates asymptotic estimation of truncation errors and analysis of convergence behavior. 
In contrast, we derive exact leading-order terms of $\nu_{\rm corr}({\mathbf r},p|{\mathbf s})$ directly in Cartesian coordinates.

Because $P_k(\cos\theta)$ is odd in ${\mathbf n}$ for odd $k$, contributions from lattice vectors $\pm {\mathbf n}$ cancel in Eq.~\eqref{eq:corrdef}; hence, only even-order multipoles ($k=2, 4, \cdots$) survive.
For instance, the $k=2$ (dipolar) contribution reads
\begin{multline} \sum_{ {\mathbf n}\notin {\cal L}(p|{\mathbf s}) }^{{\cal L}(\infty|{\mathbf s})} \frac{n^2r^2 - 3\left({\mathbf n}\cdot {\mathbf r}\right)^2} {2n^5} \\
= \sum_{ {\mathbf n}\notin {\cal L}(p|{\mathbf s}) }^{{\cal L}(\infty|{\mathbf s})} \frac{n^2r^2 - 3(x^2n_x^2+y^2n_y^2+z^2n_z^2) } {2n^5} \label{eq:k2}, \end{multline}
where the cross terms in the expansion of $\left({\mathbf n}\cdot {\mathbf r}\right)^2$ vanish upon summation because of the odd symmetry under $n_\alpha \to  - n_\alpha$.
In general, the $k$-th ($k$ even) multipole contributes a combination of terms such as $x^k$, $y^k$, $x^{k-2}y^2$, etc., with coefficients determined by the shape and size of the crystal.
Their scaling behavior can be obtained via asymptotic analysis.
For $ r \ll n$, the discrete sum in Eq.~\eqref{eq:corrdef} may be approximated by a continuous integral over the excluded volume $\Omega(\infty|{\mathbf s}) - \Omega(p|{\mathbf s})$---the region between two similarly shaped cuboids (same aspect ratios, differing only in scale)---with a lattice-point density $1/V$.
In this continuum picture, the $k$-th multipole in the integrand decays as $n^{-(k+1)}$, while the excluded volume grows as $p^3$, suggesting a naive ${\cal O}(p^{2-k})$ scaling.

Crucially, for $k=2$ (dipolar), the continuum integral over a cuboid depends only on its shape---not its absolute size [see Eq.~\eqref{eq:ibabc}].
Since $\Omega(\infty|{\mathbf s})$ and $\Omega(p|{\mathbf s})$ share identical aspect ratios, this continuum integral over the excluded volume vanishes exactly.
Thus, the leading dipolar contribution arises solely from the discreteness of the lattice and decays as ${\cal O}(p^{-2})$.
This residual dipolar term, together with the ${\cal O}(p^{-2})$ continuum contribution from the $k=4$ (quadrupolar) term, constitutes the total leading-order finite-size correction of order $p^{-2}$.
Therefore, for a general orthogonal lattice, isolating the closed-form  $\nu _{\rm b}({\mathbf r}|{\mathbf s})$ in Eq.~\eqref{eq:ibabc} from the direct sum over a finite crystal of size $p$ yields a bulk estimate whose accuracy matches that of the CS method---without introducing renormalization.

This clean separation between boundary (shape-dependent, $p$-independent) and finite-size ($p$-dependent) corrections relies fundamentally on using cuboid-shaped finite crystals aligned with the orthogonal lattice vectors.
Such domains are self-similar under uniform scaling, enabling the exact cancellation of shape-dependent integrals over the excluded volume.
In contrast, spherical or ellipsoidal crystals---typically constructed by removing unit cells near corners and edges---lack strict shape similarity between the finite and infinite reference domains.
Consequently, the dipolar integral over the excluded region no longer vanishes, entangling boundary and finite-size effects.

While a closed-form expression for the ${\cal O}(p^{-2})$ finite-size correction remains challenging for a general orthogonal lattice, it can be derived explicitly for the cubic case---where the crystal $(a=b=c)$ and lattice $(l_x=l_y=l_z)$ share cubic symmetry and ${\mathbf s}={\mathbf 0}$.
Cubic symmetry enforces
\begin{equation} \sum_{ {\mathbf n}\notin {\cal L}(p|{\mathbf 0}) }^{{\cal L}(\infty|{\mathbf 0})} \frac{ 3n_x^2}{n^5} = \sum_{ {\mathbf n}\notin {\cal L}(p|{\mathbf 0}) }^{{\cal L}(\infty|{\mathbf 0})} \frac{1}{n^3} \label{eq:cubsym}, \end{equation}
and analogously for $n_y^2$ and $n_z^2$. As a result, Eq.~\eqref{eq:k2} completely vanishes, i.e., both the continuum approximation and the residual discrete dipolar terms vanish.
Interestingly, Eq.~\eqref{eq:cubsym} also holds for spherical truncation of cubic unit cells. Hence, the leading correction in both cubic and spherical crystals with $l_x=l_y=l_z$ stems solely from the $k=4$ multipole.
In the SI, we evaluate the corresponding continuum integral explicitly for cubic crystals, obtaining the closed-form expression given in Eq.~\eqref{eq:corr}.
Such a closed-form expression for the spherical case, however, remains unknown. 
\section{Applications to various lattices}\label{sec:app}

\begin{table}[!htb]   
\caption{Madelung constants ($p=20$ or $K=41$) for representative ionic crystals: NaCl (rocksalt), ZnS (zincblende), and CaF$_2$ (fluorite).}\label{tab:3lattice}
\begin{tabular}{cccc}\hline method & \quad\quad NaCl \quad\quad           & \quad ZnS/Ca$^{2+}$ \quad          &      \quad F$^-$  \\[0.4ex] \hline
                            CS     &     \quad    1.7479628535 \quad      &       \quad 1.638065778 \quad      & \quad 0.8811756598 \quad      \\[0.5ex]
                            EC     &     \quad    1.7475645804 \quad      &       \quad 1.638055048 \quad      & \quad 0.8813373869 \quad      \\[0.5ex]
                            ref.   &     \quad    1.7475645946 \quad      &       \quad 1.638055053 \quad      & \quad 0.8813373865 \quad      \\\hline
\end{tabular} \end{table}
We have now derived both the boundary and finite-size corrections to the direct sum. 
This analysis shows that the boundary correction $\nu_{\rm b}({\mathbf r}|{\mathbf s})$ has a dipolar character and is independent of crystal size $p$, whereas the leading finite-size correction $\nu_{\rm corr}({\mathbf r}|{\mathbf s})$ scales as $p^{-2}$.
In general, this $p^{-2}$ scaling arises from both a residual dipolar contribution and a continuum quadrupolar contributions; for the cubic case, however, it stems solely from the quadrupolar term.
Notably, while the periodic part $\nu_{\rm pbc}({\mathbf r})$ is invariant under lattice translation, neither the boundary contribution $\nu_{\rm b}$, the finite-size correction $\nu_{\rm corr}$, nor the direct sum itself shares this periodicity. 
Consequently, evaluating these three terms at a periodically translated displacement might improve the accuracy of the bulk estimate, particularly for small $p$.

In practice, many ionic crystals adopt a cubic conventional unit cell; fractional coordinates for several examples are listed in Tab.~\ref{tab:unit}.
For such systems, the electrostatic potential---and hence the Madelung constant---can be constructed by a linear superposition of pairwise bulk contributions $\nu_{\rm pbc}({\mathbf r})$ over all symmetry-inequivalent displacements between ion pairs.

\begin{table}[!htb]   
\caption{Electrostatic energies of ions and a formula unit in a bulk CaTiO$_3$ crystal, where the fraction coordinates are: Ca$^{2+}$ at (0.5, 0.5, 0.5), Ti$^{4+}$ at (0, 0, 0), and O$^{2-}$ at (0.5, 0, 0), (0, 0.5, 0), and (0, 0, 0.5),
Energies are normalized by $e^2/d^2$, where $d$ is the distance between the nearest Ti$^{4+}$--O$^{2-}$ pair. }
\label{tab:catio3}
\begin{tabular}{lcc}\hline  ions      &  \quad\quad\quad\quad$p=20$\quad\quad\quad\quad      &   ref.        \\[0.4ex] \hline
                            Ca$^{2+}$ &                   -2.693604868                     &  -2.693604825   \\[0.5ex]
                            Ti$^{4+}$ &                   -12.37746806                     &  -12.37746803   \\[0.5ex]
                             O$^{2-}$ &                   -3.227954396                     &  -3.227954401   \\[0.5ex]
                            cell      &                   -24.75493611                     &  -24.75493606   \\\hline
\end{tabular} \end{table}
To compute Madelung constants for NaCl, ZnS, CaF$_2$ and CaTiO$_3$, we employ four displacement vectors.
One corresponds to the CsCl structure, given by ${\mathbf v}_1=(0.5, 0.5, 0.5)$; the remaining vectors (${\mathbf v}_2$, ${\mathbf v}_3$, and ${\mathbf v}_4$) and their associated $\nu_{\rm pbc}$ are listed in Tab.~\ref{tab:vec3}.
Using Eq.~\eqref{eq:phi}, we obtain the Madelung constants presented in Tabs.~\ref{tab:3lattice} and~\ref{tab:catio3}. These constants are calculated by summing the appropriately weighted $\nu_{\rm pbc}({\mathbf v}_j)$ terms, with each term analytically corrected via $\nu_{\rm b}$ and $\nu_{\rm corr}$.
Our computed reference values for NaCl and ZnS agree with established literature data\cite{naclseq,znsseq}.
For CaF$_2$, the Ca$^{2+}$ ion occupies an environment identical to that of either Zn$^{2+}$ or S$^{2-}$ in the ZnS structure.
After multiplying the sum of the Madelung constants for Ca$^{2+}$ and F$^-$ by $2$, the result also aligns with published values\cite{caf2seq}.
Notably, with $p=20$, we achieve nine-digit accuracy, which matches the intrinsic precision of $\nu_{\rm pbc}$ itself. This ${\cal O}(p^{-4})$ EC approach therefore surpasses the CS approach for all cubic lattices examined.

The necessity and magnitude of these corrections depend critically on the multipole character of the unit cell.
For CsCl, the unit cell possesses a large dipole moment, yielding a boundary contribution of $ \pi\left| {\mathbf v}_1\right|^2/\sqrt{3} = \sqrt{3}\pi/4 \simeq 1.36 $, which dominates the final reference value; 
the remaining finite-size correction [e.g. $-\sqrt{3}\nu_{\rm corr}({\mathbf v}_1,1|{\mathbf 0})/2\simeq -0.037$] and the direct sums (e.g. $\simeq 0.44$ at $p=1$) are relatively small.
Similarly, for the Zn$_4$S$_4$ and Ca$_2$F$_8$ unit cells in Tab.~\ref{tab:unit}, significant dipole and quadrupole moments exist, necessitating explicit corrections to the direct sums.

In contrast, the cubic (NaCl)$_4$ or CaTiO$_3$ unit cell forms an octopolar configuration with vanishing dipole and quadrupole moments:
\begin{equation} \sum_{j=1}^8 q_j x_{ij}^2 = \sum_{j=1}^8 q_j x_{ij}^4 = 0,  \end{equation}
and analogously for $y$ and $z$.
Consequently, as noted in earlier work (e.g., Refs.\cite{Wolf1992,Zhao_Hu2025,Tavernier2020}), various direct summation schemes without explicit corrections already yield accurate Madelung constants for NaCl. 
Interestingly, the CS method performs worse than these direct sums for NaCl; its correction cannot be simply ascribed to dipolar or quadrupolar contributions alone.
Because the renormalized distance [Eq.~\eqref{eq:rd}] depends on crystal size, it is difficult to approximate the leading order correction as an integral, making the finite-size effect harder to rationalize within the CS approach.

\section{Conclusions}\label{sec:con}
In summary, we have conducted a systematic analysis of the Madelung problem for finite crystals within a general orthogonal lattice. 
Our work resolves a conceptual ambiguity by cleanly separating the boundary correction from finite-size corrections in the direct summation. 
The resulting closed-form boundary correction provides an accurate method for computing Madelung constants with ${\cal O}(p^{-2})$ precision.
For cubic lattices, where explicit closed-form expressions for both the boundary and the leading finite-size corrections are available, the approach achieves ${\cal O}(p^{-4})$ accuracy.
Even with minimal supercells---corresponding to $p=1$ (i.e., $3^3$ unit cells)---practical calculations already attain accuracy on the order of $10^{-4}$, demonstrating applicability to ionic crystals such as CsCl, NaCl, ZnS, CaF$_2$ and CaTiO$_3$.

This work was supported by NSFC (Grant Nos. 22273047 and 21873037). Z.H is grateful to Andrei Zabolotskii for pointing out relevant references to Madelung constants.

Supporting Information: complete derivations of the boundary and finite-size corrections [Eqs.~\eqref{eq:corr} and~\eqref{eq:ibabc}].

\section{Appendix: Derivation of the Indefinite Integral via Euler Substitution}\label{sec:appdx}
\renewcommand{\theequation}{A\arabic{equation}}  
\setcounter{equation}{0}                          
In this appendix, we evaluate the following integral:
\begin{equation} I = \int \frac{du}{\left(u^2+1\right)\sqrt{u^2+2}}. \end{equation}
Using the Euler substitution
\begin{equation} \sqrt{u^2+2} + u = \tau ,\end{equation}
which maps the domain $u\in (-\infty, \infty)$ to $\tau \in (0, \infty)$, we obtain the relations: 
\begin{equation} \sqrt{u^2+2} = \frac{\tau}{2} + \frac{1}{\tau} = \tau \, \frac{du}{d\tau}, \end{equation} and
\begin{equation} u^2+1 = \frac{\tau^2}{4} + \frac{1}{\tau^2}. \end{equation}       
Substituting these expressions into the integral yields
\begin{equation} \begin{aligned}I & = \int d\tau \frac{4\tau}{\tau^4 + 4} = {\rm atan}\frac{\tau^2}{2} + C \\ & =  {\rm atan}\frac{\left( \sqrt{u^2+2} + u\right)^2}{2} + C \end{aligned}, \end{equation}
where $C$ is an integration constant and ${\rm atan}(\cdot)$ denotes the inverse tangent function .
Furthermore,  applying the subtraction formula
\begin{equation} {\rm atan} A - {\rm atan} B = {\rm atan}\frac{A - B}{1+AB}, \end{equation}
valid when $AB > -1$, we find
\begin{equation} {\rm atan}\frac{\left( \sqrt{u^2+2} + u\right)^2}{2} - {\rm atan}\frac{u}{\sqrt{u^2+2}}  = \frac{\pi}{4}. \end{equation}
Consequently, the integral can be expressed in the alternative form
\begin{equation} I = {\rm atan}\frac{u}{\sqrt{u^2+2}} + C^\prime, \end{equation}
where the constant $C^\prime$ differs form $C$ by $\pi/4$.
The Euler substitution and the arctangent identity presented here are also utilized to derive analytical results in the Supporting Information.
In the limit $p\to \infty$, the difference between the cuboid and cubic crystals results in an additional electrostatic potential at the origin, given by
\begin{multline} \int_1^3 \frac{du\, 8t^2}{\left(u^2+1\right)\sqrt{u^2+2}} = 8t^2\left({\rm atan}\frac{3}{\sqrt{11}} -\frac{\pi}{6}\right) \\ = 4t^2\left(\frac{\pi}{6} - {\rm atan}\frac{1}{3\sqrt{11}} \right), \label{eq:ibapp} \end{multline}
where the final expression follows directly from the difference between the two boundary terms (see Eq.~\eqref{eq:ibabc}).


\begin{thebibliography}{40}%
\makeatletter
\providecommand \@ifxundefined [1]{%
 \@ifx{#1\undefined}
}%
\providecommand \@ifnum [1]{%
 \ifnum #1\expandafter \@firstoftwo
 \else \expandafter \@secondoftwo
 \fi
}%
\providecommand \@ifx [1]{%
 \ifx #1\expandafter \@firstoftwo
 \else \expandafter \@secondoftwo
 \fi
}%
\providecommand \natexlab [1]{#1}%
\providecommand \enquote  [1]{``#1''}%
\providecommand \bibnamefont  [1]{#1}%
\providecommand \bibfnamefont [1]{#1}%
\providecommand \citenamefont [1]{#1}%
\providecommand \href@noop [0]{\@secondoftwo}%
\providecommand \href [0]{\begingroup \@sanitize@url \@href}%
\providecommand \@href[1]{\@@startlink{#1}\@@href}%
\providecommand \@@href[1]{\endgroup#1\@@endlink}%
\providecommand \@sanitize@url [0]{\catcode `\\12\catcode `\$12\catcode
  `\&12\catcode `\#12\catcode `\^12\catcode `\_12\catcode `\%12\relax}%
\providecommand \@@startlink[1]{}%
\providecommand \@@endlink[0]{}%
\providecommand \url  [0]{\begingroup\@sanitize@url \@url }%
\providecommand \@url [1]{\endgroup\@href {#1}{\urlprefix }}%
\providecommand \urlprefix  [0]{URL }%
\providecommand \Eprint [0]{\href }%
\providecommand \doibase [0]{http://dx.doi.org/}%
\providecommand \selectlanguage [0]{\@gobble}%
\providecommand \bibinfo  [0]{\@secondoftwo}%
\providecommand \bibfield  [0]{\@secondoftwo}%
\providecommand \translation [1]{[#1]}%
\providecommand \BibitemOpen [0]{}%
\providecommand \bibitemStop [0]{}%
\providecommand \bibitemNoStop [0]{.\EOS\space}%
\providecommand \EOS [0]{\spacefactor3000\relax}%
\providecommand \BibitemShut  [1]{\csname bibitem#1\endcsname}%
\let\auto@bib@innerbib\@empty
\bibitem [{\citenamefont {Madelung}(1918)}]{Madelung1918}%
  \BibitemOpen
  \bibfield  {author} {\bibinfo {author} {\bibfnamefont {E.}~\bibnamefont
  {Madelung}},\ }\bibfield  {title} {\enquote {\bibinfo {title} {Die
  electrostatische untersuchung des kristallgitters},}\ }\href@noop {}
  {\bibfield  {journal} {\bibinfo  {journal} {Phys. Z.}\ }\textbf {\bibinfo
  {volume} {19}},\ \bibinfo {pages} {524--533} (\bibinfo {year}
  {1918})}\BibitemShut {NoStop}%
\bibitem [{\citenamefont {de~Leeuw}\ \emph {et~al.}(1980)\citenamefont
  {de~Leeuw}, \citenamefont {Perram},\ and\ \citenamefont
  {Smith}}]{DeLeeuw_Smith1980}%
  \BibitemOpen
  \bibfield  {author} {\bibinfo {author} {\bibfnamefont {S.~W.}\ \bibnamefont
  {de~Leeuw}}, \bibinfo {author} {\bibfnamefont {J.~W.}\ \bibnamefont
  {Perram}}, \ and\ \bibinfo {author} {\bibfnamefont {E.~R.}\ \bibnamefont
  {Smith}},\ }\bibfield  {title} {\enquote {\bibinfo {title} {Simulation of
  electrostatic systems in periodic boundary conditions. i. lattice sums and
  dielectric constants},}\ }\href@noop {} {\bibfield  {journal} {\bibinfo
  {journal} {Proc. R. Soc. London, Ser. A Math. Phys. Sci.}\ }\textbf {\bibinfo
  {volume} {373}},\ \bibinfo {pages} {27--56} (\bibinfo {year}
  {1980})}\BibitemShut {NoStop}%
\bibitem [{\citenamefont {Smith}(1981)}]{Smith1981}%
  \BibitemOpen
  \bibfield  {author} {\bibinfo {author} {\bibfnamefont {E.~R.}\ \bibnamefont
  {Smith}},\ }\bibfield  {title} {\enquote {\bibinfo {title} {Electrostatic
  energy in ionic crystals},}\ }\href@noop {} {\bibfield  {journal} {\bibinfo
  {journal} {Proc. R. Soc. London, Ser. A Math. Phys. Sci.}\ }\textbf {\bibinfo
  {volume} {375}},\ \bibinfo {pages} {475} (\bibinfo {year}
  {1981})}\BibitemShut {NoStop}%
\bibitem [{\citenamefont {Smith}(2008)}]{Smith2008}%
  \BibitemOpen
  \bibfield  {author} {\bibinfo {author} {\bibfnamefont {E.~R.}\ \bibnamefont
  {Smith}},\ }\bibfield  {title} {\enquote {\bibinfo {title} {Electrostatic
  potentials in systems periodic in one, two, and three dimensions},}\
  }\href@noop {} {\bibfield  {journal} {\bibinfo  {journal} {J. Chem. Phys.}\
  }\textbf {\bibinfo {volume} {128}},\ \bibinfo {pages} {174104} (\bibinfo
  {year} {2008})}\BibitemShut {NoStop}%
\bibitem [{\citenamefont {Ballenegger}(2014)}]{Ballenegger2014}%
  \BibitemOpen
  \bibfield  {author} {\bibinfo {author} {\bibfnamefont {V.}~\bibnamefont
  {Ballenegger}},\ }\bibfield  {title} {\enquote {\bibinfo {title}
  {Communication: On the origin of the surface term in the ewald formula},}\
  }\href@noop {} {\bibfield  {journal} {\bibinfo  {journal} {J. Chem. Phys.}\
  }\textbf {\bibinfo {volume} {140}},\ \bibinfo {pages} {161102} (\bibinfo
  {year} {2014})}\BibitemShut {NoStop}%
\bibitem [{\citenamefont {Hu}(2014{\natexlab{a}})}]{Hu2014ib}%
  \BibitemOpen
  \bibfield  {author} {\bibinfo {author} {\bibfnamefont {Z.}~\bibnamefont
  {Hu}},\ }\bibfield  {title} {\enquote {\bibinfo {title} {Infinite boundary
  terms of ewald sums and pairwise interactions for electrostatics in bulk and
  at interfaces},}\ }\href@noop {} {\bibfield  {journal} {\bibinfo  {journal}
  {J. Chem. Theory Comput.}\ }\textbf {\bibinfo {volume} {10}},\ \bibinfo
  {pages} {5254--5264} (\bibinfo {year} {2014}{\natexlab{a}})}\BibitemShut
  {NoStop}%
\bibitem [{\citenamefont {Pan}\ \emph {et~al.}(2017)\citenamefont {Pan},
  \citenamefont {Yi},\ and\ \citenamefont {Hu}}]{Pan_Hu2017}%
  \BibitemOpen
  \bibfield  {author} {\bibinfo {author} {\bibfnamefont {C.}~\bibnamefont
  {Pan}}, \bibinfo {author} {\bibfnamefont {S.}~\bibnamefont {Yi}}, \ and\
  \bibinfo {author} {\bibfnamefont {Z.}~\bibnamefont {Hu}},\ }\bibfield
  {title} {\enquote {\bibinfo {title} {The effect of electrostatic boundaries
  in molecular simulations: symmetry matters},}\ }\href@noop {} {\bibfield
  {journal} {\bibinfo  {journal} {Phys. Chem. Chem. Phys.}\ }\textbf {\bibinfo
  {volume} {19}},\ \bibinfo {pages} {4861} (\bibinfo {year}
  {2017})}\BibitemShut {NoStop}%
\bibitem [{\citenamefont {Zhao}\ and\ \citenamefont {Hu}(2025)}]{Zhao_Hu2025}%
  \BibitemOpen
  \bibfield  {author} {\bibinfo {author} {\bibfnamefont {Y.}~\bibnamefont
  {Zhao}}\ and\ \bibinfo {author} {\bibfnamefont {Z.}~\bibnamefont {Hu}},\
  }\bibfield  {title} {\enquote {\bibinfo {title} {Infinite boundary terms and
  pairwise interactions: A unified framework for periodic coulomb systems},}\
  }\href@noop {} {\bibfield  {journal} {\bibinfo  {journal} {J. Chem. Theory
  Comput.}\ }\textbf {\bibinfo {volume} {21}},\ \bibinfo {pages} {5916--5927}
  (\bibinfo {year} {2025})}\BibitemShut {NoStop}%
\bibitem [{\citenamefont {Ewald}(1921)}]{Ewald1921}%
  \BibitemOpen
  \bibfield  {author} {\bibinfo {author} {\bibfnamefont {P.~P.}\ \bibnamefont
  {Ewald}},\ }\bibfield  {title} {\enquote {\bibinfo {title} {Evaluation of
  optical and electrostatic lattice potentials},}\ }\href@noop {} {\bibfield
  {journal} {\bibinfo  {journal} {Ann. Phys. Leipzig}\ }\textbf {\bibinfo
  {volume} {64}},\ \bibinfo {pages} {253--287} (\bibinfo {year}
  {1921})}\BibitemShut {NoStop}%
\bibitem [{\citenamefont {Nijboer}(1957)}]{Nijboer1957}%
  \BibitemOpen
  \bibfield  {author} {\bibinfo {author} {\bibfnamefont {F.~W.}\ \bibnamefont
  {Nijboer}, \bibfnamefont {B.~R. A.and~{De Wette}}},\ }\bibfield  {title}
  {\enquote {\bibinfo {title} {On the calculation of lattice sums},}\
  }\href@noop {} {\bibfield  {journal} {\bibinfo  {journal} {Physica}\ }\textbf
  {\bibinfo {volume} {23}},\ \bibinfo {pages} {309--321} (\bibinfo {year}
  {1957})}\BibitemShut {NoStop}%
\bibitem [{\citenamefont {Borwein}\ \emph {et~al.}(1985)\citenamefont
  {Borwein}, \citenamefont {Borwein},\ and\ \citenamefont
  {Taylor}}]{Borwein1985}%
  \BibitemOpen
  \bibfield  {author} {\bibinfo {author} {\bibfnamefont {D.}~\bibnamefont
  {Borwein}}, \bibinfo {author} {\bibfnamefont {J.~M.}\ \bibnamefont
  {Borwein}}, \ and\ \bibinfo {author} {\bibfnamefont {K.~F.}\ \bibnamefont
  {Taylor}},\ }\bibfield  {title} {\enquote {\bibinfo {title} {Convergence of
  lattice sums and madelung’s constant},}\ }\href@noop {} {\bibfield
  {journal} {\bibinfo  {journal} {J. Math. Phys.}\ }\textbf {\bibinfo {volume}
  {26}},\ \bibinfo {pages} {2999--3009} (\bibinfo {year} {1985})}\BibitemShut
  {NoStop}%
\bibitem [{\citenamefont {Evjen}(1932)}]{Evjen1932}%
  \BibitemOpen
  \bibfield  {author} {\bibinfo {author} {\bibfnamefont {H.~M.}\ \bibnamefont
  {Evjen}},\ }\bibfield  {title} {\enquote {\bibinfo {title} {On the stability
  of certain heteropolar crystals},}\ }\href@noop {} {\bibfield  {journal}
  {\bibinfo  {journal} {Phys. Rev.}\ }\textbf {\bibinfo {volume} {39}},\
  \bibinfo {pages} {675--687} (\bibinfo {year} {1932})}\BibitemShut {NoStop}%
\bibitem [{\citenamefont {Harrison}(2006)}]{Harrison2006}%
  \BibitemOpen
  \bibfield  {author} {\bibinfo {author} {\bibfnamefont {W.~A.}\ \bibnamefont
  {Harrison}},\ }\bibfield  {title} {\enquote {\bibinfo {title} {Simple
  calculation of madelung constants},}\ }\href@noop {} {\bibfield  {journal}
  {\bibinfo  {journal} {Phys. Rev. B}\ }\textbf {\bibinfo {volume} {73}},\
  \bibinfo {pages} {212103} (\bibinfo {year} {2006})}\BibitemShut {NoStop}%
\bibitem [{\citenamefont {Marathe}\ \emph {et~al.}(1983)\citenamefont
  {Marathe}, \citenamefont {Lauer},\ and\ \citenamefont
  {Trautwein}}]{Marathe1983}%
  \BibitemOpen
  \bibfield  {author} {\bibinfo {author} {\bibfnamefont {V.~R.}\ \bibnamefont
  {Marathe}}, \bibinfo {author} {\bibfnamefont {S.}~\bibnamefont {Lauer}}, \
  and\ \bibinfo {author} {\bibfnamefont {A.~X.}\ \bibnamefont {Trautwein}},\
  }\bibfield  {title} {\enquote {\bibinfo {title} {Electrostatic potentials
  using direct-lattice summations},}\ }\href@noop {} {\bibfield  {journal}
  {\bibinfo  {journal} {Phys. Rev. B}\ }\textbf {\bibinfo {volume} {27}},\
  \bibinfo {pages} {5162--5165} (\bibinfo {year} {1983})}\BibitemShut {NoStop}%
\bibitem [{\citenamefont {Sousa}\ \emph {et~al.}(1993)\citenamefont {Sousa},
  \citenamefont {Casanovas}, \citenamefont {Rubio},\ and\ \citenamefont
  {Illas}}]{Sousa1993}%
  \BibitemOpen
  \bibfield  {author} {\bibinfo {author} {\bibfnamefont {C.}~\bibnamefont
  {Sousa}}, \bibinfo {author} {\bibfnamefont {J.}~\bibnamefont {Casanovas}},
  \bibinfo {author} {\bibfnamefont {J.}~\bibnamefont {Rubio}}, \ and\ \bibinfo
  {author} {\bibfnamefont {F.}~\bibnamefont {Illas}},\ }\bibfield  {title}
  {\enquote {\bibinfo {title} {Madelung fields from optimized point charges for
  {ab initio} cluster model calculations on ionic systems},}\ }\href@noop {}
  {\bibfield  {journal} {\bibinfo  {journal} {J. Comput. Chem.}\ }\textbf
  {\bibinfo {volume} {14}},\ \bibinfo {pages} {680--684} (\bibinfo {year}
  {1993})}\BibitemShut {NoStop}%
\bibitem [{\citenamefont {Derenzo}\ \emph {et~al.}(2000)\citenamefont
  {Derenzo}, \citenamefont {Klintenberg},\ and\ \citenamefont
  {Weber}}]{Derenzo2000}%
  \BibitemOpen
  \bibfield  {author} {\bibinfo {author} {\bibfnamefont {S.~E.}\ \bibnamefont
  {Derenzo}}, \bibinfo {author} {\bibfnamefont {M.~K.}\ \bibnamefont
  {Klintenberg}}, \ and\ \bibinfo {author} {\bibfnamefont {M.~J.}\ \bibnamefont
  {Weber}},\ }\bibfield  {title} {\enquote {\bibinfo {title} {Determining point
  charge arrays that produce accurate ionic crystal fields for atomic cluster
  calculations},}\ }\href@noop {} {\bibfield  {journal} {\bibinfo  {journal}
  {J. Chem. Phys.}\ }\textbf {\bibinfo {volume} {112}},\ \bibinfo {pages}
  {2074--2081} (\bibinfo {year} {2000})}\BibitemShut {NoStop}%
\bibitem [{\citenamefont {Gell{\'e}}\ and\ \citenamefont
  {Lepetit}(2008)}]{Gelle2008}%
  \BibitemOpen
  \bibfield  {author} {\bibinfo {author} {\bibfnamefont {A.}~\bibnamefont
  {Gell{\'e}}}\ and\ \bibinfo {author} {\bibfnamefont {M.-B.}\ \bibnamefont
  {Lepetit}},\ }\bibfield  {title} {\enquote {\bibinfo {title} {Fast
  calculation of the electrostatic potential in ionic crystals by direct
  summation method},}\ }\href@noop {} {\bibfield  {journal} {\bibinfo
  {journal} {J. Chem. Phys.}\ }\textbf {\bibinfo {volume} {128}},\ \bibinfo
  {pages} {244716} (\bibinfo {year} {2008})}\BibitemShut {NoStop}%
\bibitem [{\citenamefont {Tavernier}\ \emph {et~al.}(2020)\citenamefont
  {Tavernier}, \citenamefont {Bendazzoli}, \citenamefont {Brumas},
  \citenamefont {Evangelisti},\ and\ \citenamefont {Berger}}]{Tavernier2020}%
  \BibitemOpen
  \bibfield  {author} {\bibinfo {author} {\bibfnamefont {N.}~\bibnamefont
  {Tavernier}}, \bibinfo {author} {\bibfnamefont {G.}~\bibnamefont
  {Bendazzoli}}, \bibinfo {author} {\bibfnamefont {V.}~\bibnamefont {Brumas}},
  \bibinfo {author} {\bibfnamefont {S.}~\bibnamefont {Evangelisti}}, \ and\
  \bibinfo {author} {\bibfnamefont {J.~A.}\ \bibnamefont {Berger}},\ }\bibfield
   {title} {\enquote {\bibinfo {title} {Clifford boundary conditions: A simple
  direct-sum evaluation of madelung constants},}\ }\href@noop {} {\bibfield
  {journal} {\bibinfo  {journal} {J. Phys. Chem. Lett.}\ }\textbf {\bibinfo
  {volume} {11}},\ \bibinfo {pages} {7090--7095} (\bibinfo {year}
  {2020})}\BibitemShut {NoStop}%
\bibitem [{\citenamefont {Tavernier}\ \emph {et~al.}(2021)\citenamefont
  {Tavernier}, \citenamefont {Bendazzoli}, \citenamefont {Brumas},
  \citenamefont {Evangelisti},\ and\ \citenamefont
  {Leininger}}]{Tavernier2021}%
  \BibitemOpen
  \bibfield  {author} {\bibinfo {author} {\bibfnamefont {N.}~\bibnamefont
  {Tavernier}}, \bibinfo {author} {\bibfnamefont {G.~L.}\ \bibnamefont
  {Bendazzoli}}, \bibinfo {author} {\bibfnamefont {V.}~\bibnamefont {Brumas}},
  \bibinfo {author} {\bibfnamefont {S.}~\bibnamefont {Evangelisti}}, \ and\
  \bibinfo {author} {\bibfnamefont {T.}~\bibnamefont {Leininger}},\ }\bibfield
  {title} {\enquote {\bibinfo {title} {Clifford boundary conditions for
  periodic systems: the {M}adelung constant of cubic crystals in 1, 2 and 3
  dimensions},}\ }\href@noop {} {\bibfield  {journal} {\bibinfo  {journal}
  {Theor. Chem. Acc.}\ }\textbf {\bibinfo {volume} {140}},\ \bibinfo {pages}
  {106} (\bibinfo {year} {2021})}\BibitemShut {NoStop}%
\bibitem [{csc()}]{csclseq}%
  \BibitemOpen
  \href@noop {} {}\bibinfo {note} {Http://oeis.org/A181152, Madelung Constant
  for CsCl Structure, The Online Encyclopedia of Integer
  Sequences.}\BibitemShut {Stop}%
\bibitem [{\citenamefont {Rayleigh}(1892)}]{Rayleigh1892}%
  \BibitemOpen
  \bibfield  {author} {\bibinfo {author} {\bibfnamefont {Lord}\ \bibnamefont
  {Rayleigh}},\ }\bibfield  {title} {\enquote {\bibinfo {title} {On the
  influence of obstacles arranged in rectangular order upon the properties of a
  medium},}\ }\href@noop {} {\bibfield  {journal} {\bibinfo  {journal} {Phil.
  Mag. Ser. 5}\ }\textbf {\bibinfo {volume} {34}},\ \bibinfo {pages} {481--502}
  (\bibinfo {year} {1892})}\BibitemShut {NoStop}%
\bibitem [{\citenamefont {de~Arag{\~a}o}\ \emph {et~al.}(2019)\citenamefont
  {de~Arag{\~a}o}, \citenamefont {Moreno}, \citenamefont {Battaglia},
  \citenamefont {Bendazzoli}, \citenamefont {Evangelisti}, \citenamefont
  {Leininger}, \citenamefont {Suaud},\ and\ \citenamefont
  {Berger}}]{Aragao2019}%
  \BibitemOpen
  \bibfield  {author} {\bibinfo {author} {\bibfnamefont {E.}~\bibnamefont
  {de~Arag{\~a}o}}, \bibinfo {author} {\bibfnamefont {D.}~\bibnamefont
  {Moreno}}, \bibinfo {author} {\bibfnamefont {S.}~\bibnamefont {Battaglia}},
  \bibinfo {author} {\bibfnamefont {G.}~\bibnamefont {Bendazzoli}}, \bibinfo
  {author} {\bibfnamefont {S.}~\bibnamefont {Evangelisti}}, \bibinfo {author}
  {\bibfnamefont {T.}~\bibnamefont {Leininger}}, \bibinfo {author}
  {\bibfnamefont {N.}~\bibnamefont {Suaud}}, \ and\ \bibinfo {author}
  {\bibfnamefont {J.~A.}\ \bibnamefont {Berger}},\ }\bibfield  {title}
  {\enquote {\bibinfo {title} {A simple position operator for periodic
  systems},}\ }\href@noop {} {\bibfield  {journal} {\bibinfo  {journal} {Phys.
  Rev. B}\ }\textbf {\bibinfo {volume} {99}},\ \bibinfo {pages} {205145}
  (\bibinfo {year} {2019})}\BibitemShut {NoStop}%
\bibitem [{\citenamefont {Hummer}(1996)}]{Hummer1996}%
  \BibitemOpen
  \bibfield  {author} {\bibinfo {author} {\bibfnamefont {Gerhard}\ \bibnamefont
  {Hummer}},\ }\bibfield  {title} {\enquote {\bibinfo {title} {Electrostatic
  potential of a homogeneously charged square and cube in two and three
  dimensions},}\ }\href@noop {} {\bibfield  {journal} {\bibinfo  {journal} {J.
  Electrostatics}\ }\textbf {\bibinfo {volume} {36}},\ \bibinfo {pages}
  {285--291} (\bibinfo {year} {1996})}\BibitemShut {NoStop}%
\bibitem [{\citenamefont {Ciftja}(2020)}]{Ciftja2020}%
  \BibitemOpen
  \bibfield  {author} {\bibinfo {author} {\bibfnamefont {Orion}\ \bibnamefont
  {Ciftja}},\ }\bibfield  {title} {\enquote {\bibinfo {title} {Electrostatic
  potential of a uniformly charged square plate at an arbitrary point in
  space},}\ }\href@noop {} {\bibfield  {journal} {\bibinfo  {journal} {Physica
  Scripta}\ }\textbf {\bibinfo {volume} {95}},\ \bibinfo {pages} {095802}
  (\bibinfo {year} {2020})}\BibitemShut {NoStop}%
\bibitem [{\citenamefont {Yi}\ \emph {et~al.}(2017)\citenamefont {Yi},
  \citenamefont {Pan},\ and\ \citenamefont {Hu}}]{Yi_Hu2017pairwise}%
  \BibitemOpen
  \bibfield  {author} {\bibinfo {author} {\bibfnamefont {S.}~\bibnamefont
  {Yi}}, \bibinfo {author} {\bibfnamefont {C.}~\bibnamefont {Pan}}, \ and\
  \bibinfo {author} {\bibfnamefont {Z.}~\bibnamefont {Hu}},\ }\bibfield
  {title} {\enquote {\bibinfo {title} {Note: A pairwise form of the ewald sum
  for non-neutral systems},}\ }\href@noop {} {\bibfield  {journal} {\bibinfo
  {journal} {J. Chem. Phys.}\ }\textbf {\bibinfo {volume} {147}},\ \bibinfo
  {pages} {126101} (\bibinfo {year} {2017})}\BibitemShut {NoStop}%
\bibitem [{\citenamefont {Hu}(2022)}]{Hu2022}%
  \BibitemOpen
  \bibfield  {author} {\bibinfo {author} {\bibfnamefont {Z.}~\bibnamefont
  {Hu}},\ }\bibfield  {title} {\enquote {\bibinfo {title} {The
  symmetry-preserving mean field condition for electrostatic correlations in
  bulk},}\ }\href@noop {} {\bibfield  {journal} {\bibinfo  {journal} {J. Chem.
  Phys.}\ }\textbf {\bibinfo {volume} {156}},\ \bibinfo {pages} {034111}
  (\bibinfo {year} {2022})}\BibitemShut {NoStop}%
\bibitem [{\citenamefont {Gao}\ \emph {et~al.}(2023)\citenamefont {Gao},
  \citenamefont {Hu},\ and\ \citenamefont {Xu}}]{Gao_Hu2023}%
  \BibitemOpen
  \bibfield  {author} {\bibinfo {author} {\bibfnamefont {W.}~\bibnamefont
  {Gao}}, \bibinfo {author} {\bibfnamefont {Z.}~\bibnamefont {Hu}}, \ and\
  \bibinfo {author} {\bibfnamefont {Z.}~\bibnamefont {Xu}},\ }\bibfield
  {title} {\enquote {\bibinfo {title} {A screening condition imposed stochastic
  approximation for long-range electrostatic correlations},}\ }\href@noop {}
  {\bibfield  {journal} {\bibinfo  {journal} {J. Chem. Theory Comput.}\
  }\textbf {\bibinfo {volume} {19}},\ \bibinfo {pages} {4822--4827} (\bibinfo
  {year} {2023})}\BibitemShut {NoStop}%
\bibitem [{\citenamefont {Hu}(2014{\natexlab{b}})}]{Hu2014spmf}%
  \BibitemOpen
  \bibfield  {author} {\bibinfo {author} {\bibfnamefont {Z.}~\bibnamefont
  {Hu}},\ }\bibfield  {title} {\enquote {\bibinfo {title} {A
  symmetry-preserving mean field theory for electrostatics at interfaces},}\
  }\href@noop {} {\bibfield  {journal} {\bibinfo  {journal} {Chem. Commun.}\
  }\textbf {\bibinfo {volume} {50}},\ \bibinfo {pages} {14397--14400} (\bibinfo
  {year} {2014}{\natexlab{b}})}\BibitemShut {NoStop}%
\bibitem [{\citenamefont {Pan}\ \emph {et~al.}(2019)\citenamefont {Pan},
  \citenamefont {Yi},\ and\ \citenamefont {Hu}}]{Pan_Hu2019}%
  \BibitemOpen
  \bibfield  {author} {\bibinfo {author} {\bibfnamefont {C.}~\bibnamefont
  {Pan}}, \bibinfo {author} {\bibfnamefont {S.}~\bibnamefont {Yi}}, \ and\
  \bibinfo {author} {\bibfnamefont {Z.}~\bibnamefont {Hu}},\ }\bibfield
  {title} {\enquote {\bibinfo {title} {Analytic theory of finite-size effects
  in supercell modelling of charged interfaces},}\ }\href@noop {} {\bibfield
  {journal} {\bibinfo  {journal} {Phys. Chem. Chem. Phys.}\ }\textbf {\bibinfo
  {volume} {21}},\ \bibinfo {pages} {14858} (\bibinfo {year}
  {2019})}\BibitemShut {NoStop}%
\bibitem [{\citenamefont {Onegin}\ \emph {et~al.}(2024)\citenamefont {Onegin},
  \citenamefont {Demyanov},\ and\ \citenamefont {Levashov}}]{Onegin2024}%
  \BibitemOpen
  \bibfield  {author} {\bibinfo {author} {\bibfnamefont {A~S}\ \bibnamefont
  {Onegin}}, \bibinfo {author} {\bibfnamefont {G~S}\ \bibnamefont {Demyanov}},
  \ and\ \bibinfo {author} {\bibfnamefont {P~R}\ \bibnamefont {Levashov}},\
  }\bibfield  {title} {\enquote {\bibinfo {title} {Pressure of coulomb systems
  with volume-dependent long-range potentials},}\ }\href@noop {} {\bibfield
  {journal} {\bibinfo  {journal} {J. Phys. A: Math. Theor.}\ }\textbf {\bibinfo
  {volume} {57}},\ \bibinfo {pages} {205002} (\bibinfo {year}
  {2024})}\BibitemShut {NoStop}%
\bibitem [{\citenamefont {Li}\ \emph {et~al.}(2025)\citenamefont {Li},
  \citenamefont {Liang},\ and\ \citenamefont {Xu}}]{Li2025}%
  \BibitemOpen
  \bibfield  {author} {\bibinfo {author} {\bibfnamefont {Lei}\ \bibnamefont
  {Li}}, \bibinfo {author} {\bibfnamefont {Jiuyang}\ \bibnamefont {Liang}}, \
  and\ \bibinfo {author} {\bibfnamefont {Zhenli}\ \bibnamefont {Xu}},\
  }\bibfield  {title} {\enquote {\bibinfo {title} {Comment on ‘pressure of
  coulomb systems with volume-dependent long-range potentials’},}\
  }\href@noop {} {\bibfield  {journal} {\bibinfo  {journal} {J. Phys. A: Math.
  Theor.}\ }\textbf {\bibinfo {volume} {58}},\ \bibinfo {pages} {088001}
  (\bibinfo {year} {2025})}\BibitemShut {NoStop}%
\bibitem [{\citenamefont {Demyanov}\ \emph {et~al.}(2025)\citenamefont
  {Demyanov}, \citenamefont {Onegin},\ and\ \citenamefont
  {Levashov}}]{Demyanov2025}%
  \BibitemOpen
  \bibfield  {author} {\bibinfo {author} {\bibfnamefont {G~S}\ \bibnamefont
  {Demyanov}}, \bibinfo {author} {\bibfnamefont {A~S}\ \bibnamefont {Onegin}},
  \ and\ \bibinfo {author} {\bibfnamefont {P~R}\ \bibnamefont {Levashov}},\
  }\bibfield  {title} {\enquote {\bibinfo {title} {Reply to comment on
  ‘pressure of coulomb systems with volume-dependent long-range
  potentials’},}\ }\href@noop {} {\bibfield  {journal} {\bibinfo  {journal}
  {J. Phys. A: Math. Theor.}\ }\textbf {\bibinfo {volume} {58}},\ \bibinfo
  {pages} {088002} (\bibinfo {year} {2025})}\BibitemShut {NoStop}%
\bibitem [{\citenamefont {Pan}(2017)}]{Pan2017}%
  \BibitemOpen
  \bibfield  {author} {\bibinfo {author} {\bibfnamefont {C.}~\bibnamefont
  {Pan}},\ }\emph {\bibinfo {title} {A study on electrostatics algorithms in
  molecular simulations}},\ \href@noop {} {Ph.D. thesis},\ \bibinfo  {school}
  {Jilin University} (\bibinfo {year} {2017}),\ \bibinfo {note} {pages
  139-140}\BibitemShut {NoStop}%
\bibitem [{\citenamefont {Greengard}\ and\ \citenamefont
  {Rokhlin}(1987)}]{Greengard1987}%
  \BibitemOpen
  \bibfield  {author} {\bibinfo {author} {\bibfnamefont {Leslie}\ \bibnamefont
  {Greengard}}\ and\ \bibinfo {author} {\bibfnamefont {Vladimir}\ \bibnamefont
  {Rokhlin}},\ }\bibfield  {title} {\enquote {\bibinfo {title} {A fast
  algorithm for particle simulations},}\ }\href@noop {} {\bibfield  {journal}
  {\bibinfo  {journal} {J. Comput. Phys.}\ }\textbf {\bibinfo {volume} {73}},\
  \bibinfo {pages} {325--348} (\bibinfo {year} {1987})}\BibitemShut {NoStop}%
\bibitem [{\citenamefont {Greengard}(1988)}]{Greengard1988}%
  \BibitemOpen
  \bibfield  {author} {\bibinfo {author} {\bibfnamefont {Leslie}\ \bibnamefont
  {Greengard}},\ }\href@noop {} {\emph {\bibinfo {title} {The Rapid Evaluation
  of Potential Fields in Particle Systems}}}\ (\bibinfo  {publisher} {MIT
  Press},\ \bibinfo {address} {Cambridge, MA},\ \bibinfo {year}
  {1988})\BibitemShut {NoStop}%
\bibitem [{\citenamefont {Kudin}\ and\ \citenamefont
  {Scuseria}(1998)}]{Kudin1998}%
  \BibitemOpen
  \bibfield  {author} {\bibinfo {author} {\bibfnamefont {Konstantin~N}\
  \bibnamefont {Kudin}}\ and\ \bibinfo {author} {\bibfnamefont {Gustavo~E}\
  \bibnamefont {Scuseria}},\ }\bibfield  {title} {\enquote {\bibinfo {title} {A
  fast multipole method for periodic systems with arbitrary unit cell
  geometries},}\ }\href@noop {} {\bibfield  {journal} {\bibinfo  {journal}
  {Chem. Phys. Lett.}\ }\textbf {\bibinfo {volume} {283}},\ \bibinfo {pages}
  {61--68} (\bibinfo {year} {1998})}\BibitemShut {NoStop}%
\bibitem [{nac()}]{naclseq}%
  \BibitemOpen
  \href@noop {} {}\bibinfo {note} {Http://oeis.org/A085469, Madelung Constant
  for NaCl Structure, The Online Encyclopedia of Integer
  Sequences.}\BibitemShut {Stop}%
\bibitem [{zns()}]{znsseq}%
  \BibitemOpen
  \href@noop {} {}\bibinfo {note} {Http://oeis.org/A182566, Madelung Constant
  for ZnS Structure, The Online Encyclopedia of Integer Sequences.}\BibitemShut
  {Stop}%
\bibitem [{caf()}]{caf2seq}%
  \BibitemOpen
  \href@noop {} {}\bibinfo {note} {Http://oeis.org/A182567, Madelung Constant
  for CaF$_2$ Structure, The Online Encyclopedia of Integer
  Sequences.}\BibitemShut {Stop}%
\bibitem [{\citenamefont {Wolf}(1992)}]{Wolf1992}%
  \BibitemOpen
  \bibfield  {author} {\bibinfo {author} {\bibfnamefont {D.}~\bibnamefont
  {Wolf}},\ }\bibfield  {title} {\enquote {\bibinfo {title} {Reconstruction of
  nacl surfaces from a dipolar solution to the madelung problem},}\ }\href@noop
  {} {\bibfield  {journal} {\bibinfo  {journal} {Phys. Rev. Lett.}\ }\textbf
  {\bibinfo {volume} {68}},\ \bibinfo {pages} {3315--3318} (\bibinfo {year}
  {1992})}\BibitemShut {NoStop}%
\end{thebibliography}

%
\section*{Supporting information}\label{sec:si}
\renewcommand{\theequation}{S\arabic{equation}}  
\setcounter{equation}{0}                          
This Supporting Information presents explicit evaluations of the two integrals over centro-symmetric domains: a general cuboid $\Omega(a,b,c)=[-a/2,a/2]\times[-b/2,b/2]\times[-c/2,c/2]$ and a cube $\Omega(a,a,a)=[-a/2,a/2]^3$. 
Up to a multiplicative factor of $V=l_xl_yl_z$ (the volume of the unit cell), these integrals correspond to the infinite boundary term and the quadrupolar correction, respectively.
The results are as follows:
\begin{multline} I_1({\mathbf r},a,b,c) =  \frac{1}{2}\int_{\Omega(a,b,c)} d{\mathbf x} \, \left({\mathbf r}\cdot \nabla_{\mathbf x}\right)^2\frac{1}{\left| {\mathbf x}\right|}  \\ = -4\sum_{\alpha=1}^3 ({\mathbf r}\cdot {\mathbf e}_\alpha)^2 {\rm atan}\frac{1}{\xi_\alpha^2\sqrt{\xi_1^2+\xi_2^2+\xi_3^2}}, \label{eq:ibsq} \end{multline}
and
\begin{equation} \begin{aligned} I_2({\mathbf r},a) &= \frac{1}{24}\int_{[-a/2,a/2]^3} d{\mathbf x} \, \left({\mathbf r}\cdot \nabla_{\mathbf x}\right)^4\frac{1}{\left| {\mathbf x}\right|} \\ &=  \frac{24r^4 - 40(x^4+y^4+z^4)}{9\sqrt{3} a^2} \end{aligned}  \label{eq:i4}, \end{equation}
where ${\mathbf x}=(x_1,x_2,x_3)$ is the integration variable, and ${\mathbf r}=(x,y,z)$ is a given vector with $r^2=x^2+y^2+z^2$.
In Eq.~\eqref{eq:ibsq}, ${\mathbf e}_1$, ${\mathbf e}_2$, and ${\mathbf e}_3$ denote the orthogonal unit vectors along the $x$, $y$, $z$ directions, respectively, so that ${\mathbf r}\cdot{\mathbf e}_1 = x$, etc. 
The dimensionless parameters $\xi_\alpha$ characterize the shape of the cuboid and are defined by normalizing each side length with the geometric mean $(abc)^{1/3}$:
\begin{equation}\xi_1 = \frac{a}{(abc)^{1/3}}; \quad \xi_2 = \frac{b}{(abc)^{1/3}};\quad \xi_3 = \frac{c}{(abc)^{1/3}}, \end{equation}
which satisfy $\xi_1\xi_2\xi_3 = 1$. These parameters depend only on the relative proportions $a:b:c$, not the overall size of the cuboid.
The function ${\rm atan}(\cdot)$ in Eq.~\eqref{eq:ibsq} denotes the inverse tangent (arctangent) and obeys the subtraction identity: 
\begin{equation} {\rm atan}(A) - {\rm atan}(B) = {\rm atan}\frac{A-B}{1+AB}. \label{eq:atansub}\end{equation}

One of the results ($I_1$) was previously obtained by Pan\cite{Pan2017} and is closely related to Eq.(3.26) in the work by Smith\cite{Smith1981}; we have not found any evaluation of $I_2$ in the literature.
\subsection{Preliminary Integrals}
Before proceeding to the derivations, we list several preliminary definite integrals used throughout the evaluations, valid for $x, t, \tau >0$:
\begin{equation} \int_0^x \frac{du\,\tau}{u^2+\tau^2} = {\rm atan}(x/\tau), \end{equation}
\begin{equation} \int_0^x \frac{du\,(u^2+t^2)^{-1}}{\sqrt{u^2+t^2+\tau^2}} = \frac{1}{t\tau}{\rm atan}\frac{\tau x}{t\sqrt{x^2+t^2+\tau^2}}, \label{eq:euler} \end{equation}
\begin{equation}\int_0^x \frac{du}{(u^2+\tau^2)^{3/2}} = \frac{x}{\tau^2 \sqrt{ \tau^2 + x^2 }},  \label{eq:i3} \end{equation}

\begin{equation}\int_0^x \frac{du}{(u^2+\tau^2)^{5/2}} = \frac{3\tau^2x + 2x^3}{3\tau^4( \tau^2 + x^2 )^{3/2}}, \label{eq:i5}  \end{equation}
\begin{equation}\int_0^x \frac{du}{(u^2+\tau^2)^{7/2}} = \frac{15\tau^4x + 20\tau^2x^3+8x^5}{15\tau^6( \tau^2 + x^2 )^{5/2}}, \label{eq:i7}  \end{equation}
\begin{equation}\int_0^x \frac{du\, u^2}{(u^2+\tau^2)^{7/2}} = \frac{5\tau^2x^3 + 2x^5}{15\tau^4( \tau^2 + x^2 )^{5/2}}, \label{eq:i27}  \end{equation}
\begin{multline}\int_0^x \frac{du\, (11+7u^2)}{(1+u^2)^3 (2+u^2)^{5/2}} =\\ \frac{x(2x^6+10x^4+17x^2+11)}{2(x^2+1)^2(x^2+2)^{3/2}} \label{eq:i35}. \end{multline}
Eq.~\eqref{eq:euler} can be derived via the Euler substitution $w = u+\sqrt{u^2+t^2+\tau^2}$ which rationalizes the square root and reduces the integral to a standard arctangent form.
Eqs.~\eqref{eq:i3} to~\eqref{eq:i35} were obtained via an ansatz method: assuming an antiderivative of the form $P(u)/Q(u)$, where $Q(u)$ matches the expected denominator structure, then determining the coefficients of the polynomial $P(u)$ by differentiation and matching.
We also list a number of definite integrals involving the error function defined as
\begin{equation} {\rm erf}(\tau) = \frac{2}{\sqrt{\pi}} \int_0^\tau du\, e^{-u^2}  \label{eq:erf}. \end{equation}
\begin{equation} \int_{-\infty}^{\infty} du\,   e^{-tu^2}  \frac{\sin(u\tau)}{u} = \pi {\rm erf}\left(\frac{\tau}{2\sqrt{t}} \right),  \label{eq:esin} \end{equation}
\begin{equation} \int_0^\infty du\, \frac{ {\rm erf}(u \tau)  u}{e^{u^2}} = \frac{1}{2} \frac{\tau}{\sqrt{1+\tau^2}}, \label{eq:ierf1} \end{equation}
\begin{equation} \int_0^\infty du\, \frac{ {\rm erf}(ut){\rm erf}(u\tau) }{e^{u^2}}  =\\ \frac{1}{\sqrt{\pi}} {\rm atan}\frac{t\tau}{\sqrt{1+\tau^2+t^2}}, \label{eq:ierf2} \end{equation}
\begin{equation} \int_{-\infty}^{\infty} du\,   e^{-tu^2} \sin(u\tau) u  = \frac{\sqrt{\pi} \tau }{2 \sqrt{t^3}} e^{-\tau^2/(4t)} \label{eq:uesin}. \end{equation}
Eqs.~\eqref{eq:esin} to~\eqref{eq:ierf2} can be derived by parametric differentiation: differentiating with respect to $\tau$ yields simpler expressions, which are integrated and then antidifferentiated back to recover the full result.
Eq.~\eqref{eq:uesin} follows from differentiating Eq.~\eqref{eq:esin} with respect to $t$.

\subsection{Derivation of $I_1$ via the divergence theorem}
To evaluate the volume integral in Eq.~\eqref{eq:ibsq}, we apply the divergence theorem (also known as Gauss's theorem). First, observe the identity
\begin{equation}\begin{aligned} \left({\mathbf r}\cdot \nabla_{\mathbf x}\right)^2\left| {\mathbf x}\right|^{-1} & = \nabla_{\mathbf x}\cdot \left[{\mathbf r} \left( {\mathbf r}\cdot \nabla_{\mathbf x}\left| {\mathbf x}\right|^{-1}\right) \right]  \\
 & =- \nabla_{\mathbf x}\cdot \left[{\mathbf r} \left({\mathbf r}\cdot {\mathbf x}\right) \left|{\mathbf x} \right|^{-3} \right], \end{aligned} \end{equation}
since $ \nabla_{\mathbf x}\left| {\mathbf x}\right|^{-1} = -{\mathbf x}/\left|{\mathbf x} \right|^3$.
Using this identity, the volume integral can be transformed into a surface integral over the boundary $\partial\Omega(a,b,c)$ of the cuboid:
\begin{equation} I_1 = -\frac{1}{2} \oint_{\partial\Omega(a,b,c)} {\mathbf r}\cdot{d\mathbf S} \, \frac{{\mathbf r}\cdot {\mathbf x}}{\left| {\mathbf x} \right|^3}, \label{eq:oint} \end{equation}
where $d{\mathbf S}$ denotes an infinitesimal vector element of surface area, directed outward from the enclosed volume.

The cuboid has six faces---three pairs of opposite rectangles lying in planes parallel to the $xy$, $yz$, and $zx$ coordinate planes, respectively. 
Considering the face at $x_3 = +c/2$ whose outward normal is $+{\mathbf e}_3$. On this face, $d{\mathbf S}=dx_1 dx_2 {\mathbf e}_3$ and ${\mathbf x}=(x_1,x_2,c/2)$, so
\begin{equation} {\mathbf r}\cdot d{\mathbf S} \left({\mathbf r}\cdot{\mathbf x}\right) =  dx_1 dx_2\, \left( xz\, x_1 + yz\, x_2 + z^2c/2 \right). \end{equation}
Due to the even symmetry of the integration domain $[-a/2,a/2]\times[-b/2,b/2]$ in both $x_1$ and $x_2$, the terms linear in $x_1$ or $x_2$ integrate to zero. Only the term proportional to $z^2$ survives.
On the opposite face at $x_3 = -c/2$, the outward normal is $-{\mathbf e}_3$, giving ${\mathbf r}\cdot d{\mathbf S} \left({\mathbf r}\cdot{\mathbf x}\right) =  dx_1 dx_2\, \left( -xz x_1 - yz x_2 + z^2c/2 \right)$ and again yielding a non-zero contribution only from the $z^2$ term upon integration.  
Thus, the total contribution from the pair of faces ($x_3 = \pm c/2$) doubles the surviving quadratic term. Analogous reasoning applies to the other two pairs of faces. Hence, the full integral becomes
\begin{equation} I_1  =  z^2 S_{xy} + x^2 S_{yz} + y^2 S_{zx}  \label{eq:isxyz}, \end{equation}
where $S_{xy}$, $S_{yz}$, and $S_{zx}$ denote the coefficients of the quadratic terms contributed by the pair of faces parallel to the $xy$-, $yz$-, and $zx$-planes, respectively.
Explicitly,
\begin{equation} S_{xy} = \int_{-a/2}^{a/2} dx_1 \int_{-b/2}^{b/2} dx_2 \, \frac{-c/2}{\left( x_1^2 + x_2^2  + c^2/4\right)^{3/2}}.  \end{equation}
Exploiting evenness in $x_1$ and $x_2$ and using the substitutions $u=2x_1$ and $v=2x_2$, we obtain
\begin{equation} \begin{aligned}  S_{xy} &= -4 c \int_0^a du\int_0^b dv\, \frac{1}{ \left( u^2 + v^2  + c^2\right)^{3/2}} \\ &= -4 bc \int_0^a du\, \frac{1}{u^2+c^2}\frac{1}{\sqrt{u^2+b^2+c^2}}, \end{aligned} \end{equation}
where the integration over $dv$ is of the form in Eq.~\eqref{eq:i3}. Applying Eq.~\eqref{eq:euler} to the remaining integral yields
\begin{equation} \begin{aligned} S_{xy} &= -4 {\rm atan}\frac{ab}{c\sqrt{a^2+b^2+c^2}} \\ & = -4 {\rm atan}\frac{1}{\xi_3^2\sqrt{\xi_1^2+\xi_2^2+\xi_3^2}}. \end{aligned} \end{equation}
By symmetry, the other two pairs of faces contribute
\begin{equation} S_{yz}=-4{\rm atan}\frac{1}{\xi_1^2\sqrt{\xi_1^2+\xi_2^2+\xi_3^2}}, \label{eq:syz} \end{equation}
and
\begin{equation} S_{zx}=-4{\rm atan}\frac{1}{\xi_2^2\sqrt{\xi_1^2+\xi_2^2+\xi_3^2}}, \end{equation}
respectively. 
Summing all contributions yields the desired result in Eq.~\eqref{eq:ibsq}. Using the subtraction identity in Eq.~\eqref{eq:atansub} and noting that $\xi_1\xi_2\xi_3=1$, one finds,
\begin{equation} S_{xy} + S_{yz} + S_{zx} = - 2\pi .\end{equation}
\subsection {Derivation of $I_1$ via the Fourier transform}
Instead of employing the divergence theorem, the integral
\begin{equation} I_1({\mathbf r},2a,2b,2c) = \frac{1}{2}\int_{\Omega(2a,2b,2c)}d{\mathbf x} \left({\mathbf r}\cdot \nabla_{\mathbf x}\right)^2\frac{1}{\left| {\mathbf x}\right|} \end{equation}
can be evaluated using Fourier transform techniques. Here, $\Omega(2a,2b,2c)=[-a,a]\times[-b,b]\times[-c,c]$ denotes a cuboid of side lengths $2a$, $2b$, and $2c$, which shares the same aspect ratio (i.e., shape) as $\Omega(a,b,c)$. 
As shown below, the integration yields the same result as in Eq.~\eqref{eq:ibsq}. 
We begin with the inverse Fourier transform:
\begin{equation}  \frac{1}{\left| {\mathbf x} \right|} = \frac{1}{8\pi^3}\int_{{\mathbb R}^3} d{\mathbf k}\, e^{i{\mathbf k}\cdot{\mathbf x}} \frac{4\pi}{k^2}, \end{equation}
where ${\mathbf k}=(k_1,k_2,k_3)$ and $k^2=\left| {\mathbf k} \right|^2=k_1^2+k_2^2+k_3^2$.
Taking two derivatives gives
\begin{equation}    \left({\mathbf r}\cdot \nabla_{\mathbf x}\right)^2\frac{1}{\left| {\mathbf x}\right|}   = -\frac{1}{2\pi^2} \int_{{\mathbb R}^3} d{\mathbf k}\, e^{i{\mathbf k}\cdot{\mathbf x}}  \frac{\left({\mathbf r}\cdot{\mathbf k}\right)^2}{k^2},         \end{equation}
and then interchanging the order of integrations yields
\begin{equation} I_1 = -\frac{1}{4\pi^2} \int_{{\mathbb R}^3} d{\mathbf k}\, \frac{\left({\mathbf r}\cdot{\mathbf k}\right)^2}{k^2} \left[ \int_{\Omega(2a,2b,2c)}d{\mathbf x}\,  e^{i{\mathbf k}\cdot{\mathbf x}}   \right]  \label{eq:i1prime}.  \end{equation}
The inner integral factorizes:
\begin{equation} \int_{\Omega(2a,2b,2c)}d{\mathbf x} \, e^{i{\mathbf k}\cdot{\mathbf x}} = \frac{\sin(k_1a)}{k_1/2} \frac{\sin(k_2b)}{k_2/2} \frac{\sin(k_3c)}{k_3/2},  \label{eq:xcub} \end{equation}
which is even in $k_1$, $k_2$ and $k_3$.
To handle the $1/k^2$ term, we use the identity from a Gaussian integral:
\begin{equation} \frac{1}{k^2} = \int_0^\infty dw\, e^{-k^2 w} = \int_0^\infty dw\, e^{-k_1^2 w} e^{-k_2^2 w} e^{-k_3^2 w}. \label{eq:igk2}  \end{equation}
Substituting Eqs.~\eqref{eq:xcub} and~\eqref{eq:igk2} into Eq.~\eqref{eq:i1prime} allows us to write $I_1$ as a four-dimensional integral, three over $d{\mathbf k}=dk_1 dk_2 dk_3$ and one over $dw$. 
Expanding $\left({\mathbf r}\cdot{\mathbf k}\right)^2 = x^2k_1^2 + y^2k_2^2 + z^2k_3^2 + \text{cross terms}$, we observe that all cross terms (e.g. $k_1k_2$) integrate to zero due to odd symmetry under $k_1 \to -k_1$ etc. Thus,
\begin{equation} I_1 = x^2 C_x + y^2 C_y + z^2 C_z \label{eq:icxyz}, \end{equation}
where each coefficient arises from integrating one quadratic component. The integrals over $d{\mathbf k}$ factorize and can be evaluated using Eqs~\eqref{eq:esin} and~\eqref{eq:uesin}. For example,  for the $x^2$-term:
\begin{equation} C_x = \int_0^\infty \frac{dw\,  a}{\sqrt{w^3} }\frac{-\sqrt{\pi} }{e^{a^2/(4w)} }        {\rm erf}\left(\frac{b/2}{\sqrt{w}}\right)     {\rm erf}\left(\frac{c/2}{\sqrt{w}} \right) . \end{equation}
Now applying the substitution $ 2\sqrt{w} = a/u $, so that
\begin{equation} \frac{dw}{\sqrt{w^3}} = 4\frac{du}{a}; \quad \frac{a^2}{4w} = u^2;\quad \frac{b}{2\sqrt{w}} = \frac{b}{a} u;\quad \text{etc.} \end{equation}
and
\begin{equation} C_x = - 4\sqrt{\pi} \int_0^\infty du\, e^{-u^2} {\rm erf}\left(u\frac{b}{a}\right) {\rm erf}\left(u\frac{c}{a}\right).  \label{eq:smith} \end{equation}
Under this change of variables, the integral in $C_x$ now matches the form of Eq.~\eqref{eq:ierf2} with $t=b/a$ and $\tau = c/a$, yielding $C_x = S_{yz}$ (see Eq.~\eqref{eq:syz}).
By symmetry, $C_y$ and $C_z$ are obtained analogously (e.g., exchanging $a \leftrightarrow b$ in $C_x$ gives $C_y = S_{zx}$, and $a \leftrightarrow c$ gives $C_z = S_{xy}$). 
Thus, Eq.~\eqref{eq:icxyz} recovers the result in Eq.~\eqref{eq:isxyz} or~\eqref{eq:ibsq}.

This alternative derivation has certain advantages. For example, from Eqs.~\eqref{eq:i1prime},~\eqref{eq:xcub} and the Dirichlet integral
\begin{equation} \int_{-\infty}^\infty du\, \frac{\sin(u\tau)}{u}=\pi ;\quad \text{for}\quad \tau > 0, \end{equation} 
it follows that the sum of the coefficients is shape-independent:
\begin{equation} C_x + C_y + C_z = - 2\pi. \end{equation}
This property is not immediately apparent in the previous derivation via the divergence theorem.

For the purpose of deriving the shape-dependent term for a macroscopic rectangular prism, Smith arrived at his Eq.(3.26) in Ref.\cite{Smith1981}, which equals the negative of Eq.~\eqref{eq:smith}. 
He wrote, ``in spite of some diligent work, no analytic progress has been made with (3.26).''
The closed-form expressions for the integrals (Eqs.~\eqref{eq:ierf1} and~\eqref{eq:ierf2} obtained by parametric differentiation) thus resolve this analytic challenge.
The connection between the shape-dependent term and $I_1({\mathbf r})$ is also transparent.
The usual shape-dependent terms are always non-negative\cite{DeLeeuw_Smith1980,Smith1981}. The one obtained by Smith is written in terms of the total dipole moment\cite{Smith1981}:
\begin{equation} J({\mathbf M}) = - \frac{1}{V} \left( C_x M_x^2 + C_y M_y^2 + C_z M_z^2 \right) ,\end{equation}
where $V = l_x l_y l_z$ is the volume of the unit cell, and the total dipole moment for a unit cell containing $N$ charges is defined as
\begin{equation} {\mathbf M} = \left(M_x,M_y,M_z\right) = \sum_{j=1}^N q_j {\mathbf r}_j .  \end{equation}
The infinite boundary term $\nu_{\rm b}(\mathbf r) = I_1({\mathbf r})/V$ is directly related to $J({\mathbf M})$ via the following pairwise sum\cite{Hu2014ib,Yi_Hu2017pairwise}
\begin{equation} \sum_{i<j}^N q_iq_j \nu_{\rm b}({\mathbf r}_i - {\mathbf r}_j) = J({\mathbf M}) - \frac{q_{\rm tot}}{V}\sum_{j=1}^N q_j |{\mathbf r}_j|^2 ,\end{equation}
where $q_{\rm tot} = \sum_{j=1}^N q_j$ is the total charge. For a neutral system ($q_{\rm tot}=0$), the second term vanishes, and the pairwise sum reduces exactly to $J({\mathbf M})$. 
\subsection{Derivation of $I_2$ via the divergence theorem}
To derive the integral
\begin{equation} I_2({\mathbf r},a) = \frac{1}{24} \int_{\Omega(a,a,a)} d{\mathbf x} \, \left({\mathbf r}\cdot \nabla_{\mathbf x}\right)^4\frac{1}{\left| {\mathbf x}\right|}, \end{equation}
we first rescale the coordinates via ${\mathbf x} \to {\mathbf x} a/2$, mapping the original cube onto the centro-symmetric domain $\Omega(2,2,2)=[-1,1]^3$ of side length 2.
After scaling, the integral simplifies to
\begin{equation} I_2({\mathbf r},a) = \frac{4}{a^2} I_2({\mathbf r},2) = \frac{1}{a^2} I_0({\mathbf r}), \end{equation}
where 
\begin{equation} I_0({\mathbf r}) = \frac{1}{6} \int_{\Omega(2,2,2)} d{\mathbf x} \, \left({\mathbf r}\cdot \nabla_{\mathbf x}\right)^4\frac{1}{\left| {\mathbf x}\right|}.  \end{equation}
This $a^{-2}$ scaling implies $I_2({\mathbf r},a) \to 0$ as $a \to \infty$, confirming that $I_2$ constitutes a finite-size correction to the bulk (infinite-domain) limit.
Below we derive the explicit form
\begin{equation} I_0({\mathbf r}) = \frac{ 24r^4 - 40(x^4+y^4+z^4) }{ 9 \sqrt{3} }, \label{eq:inti0}\end{equation}
in analogy with the derivation of $I_1$, using the divergence theorem. 

We employ the identity
\begin{equation} \left( {\mathbf r}\cdot \nabla_{\mathbf x} \right)^4 \frac{1}{|{\mathbf x}|} = 3\nabla_{\mathbf x}\cdot \left[\frac{3r^2({\mathbf r}\cdot{\mathbf x})}{|{\mathbf x}|^5}  {\mathbf r}- \frac{5({\mathbf r}\cdot{\mathbf x})^3 }{|{\mathbf x}|^7} {\mathbf r}\right] ,\end{equation}
and apply the theorem to obtain
\begin{equation} I_0 =  \frac{1}{2} \oint_{\partial\Omega(2,2,2)} {\mathbf r}\cdot d{\mathbf S}\, \left[ \frac{3r^2({\mathbf r}\cdot{\mathbf x})}{|{\mathbf x}|^5} - \frac{5({\mathbf r}\cdot{\mathbf x})^3}{|{\mathbf x}|^7} \right] .    \label{eq:oint2}      \end{equation}
Consider the face at $x_3 = +1$ (normal $+{\mathbf e}_3$). On this face:
\begin{equation} {\mathbf r}\cdot d{\mathbf S} = dx_1dx_2 \, z; \quad\quad {\mathbf x}=(x_1,x_2,+1), \end{equation}
and the integration domain for $dx_1dx_2$ is the square $[-1,1]^2$, symmetric under $x_1 \to -x_1$ and $x_2 \to -x_2$.
In the expansion of ${\mathbf r}\cdot {\mathbf x} = xx_1 +yx_2 + z$, terms odd in $x_1$ or $x_2$ vanish upon integration.
Thus only the constant term $z$ survives in ${\mathbf r}\cdot {\mathbf x}$.
For $\left( {\mathbf r}\cdot {\mathbf x}\right)^3$, we have
\begin{equation} \left({\mathbf r}\cdot{\mathbf x}\right)^3 =  z^3 + 3z (x^2 x_1^2  + y^2 x_2^2 ) + \text{odd terms}. \end{equation}
Furthermore, the square domain possesses exchange symmetry between $x_1$ and $x_2$. 
Using this symmetry and $x^2+y^2=r^2-z^2$, the combination $x^2 x_1^2  + y^2 x_2^2$ may be replaced---upon integration---by $(r^2-z^2)x_2^2$ (or equivalently by $(r^2-z^2)x_1^2$).
Thus, up to terms that integrate to zero,
\begin{equation} \left({\mathbf r}\cdot{\mathbf x}\right)^3  \to z^3 + 3z \left(r^2-z^2\right) x_2^2 .\end{equation}
By symmetry, the contributions from opposite faces are equal. 
Summing the contributions from all six faces, we write
\begin{equation} I_0 =  D_{xy} + D_{yz} + D_{zx}, \label{eq:i0d} \end{equation}
where, for instance, $D_{xy}$ denotes the combined contribution of the two faces parallel to the $xy$-plane ($x_3=\pm 1$).
Explicitly, 
\begin{equation} D_{xy} = 4r^2z^2 \left(J_5 - K_2\right) -  4z^4\left( J_7 -  K_2 \right)  , \label{eq:dc} \end{equation}
with the auxiliary integrals
\begin{equation} J_5 = \int_0^1 dx_1\int_0^1dx_2\,\frac{3}{\left(x_1^2 + x_2^2 + 1\right)^{5/2}}, \end{equation}
\begin{equation} J_7 = \int_0^1 dx_1\int_0^1dx_2\,\frac{5}{\left(x_1^2 + x_2^2 + 1\right)^{7/2}}, \end{equation}
and 
\begin{equation} K_2 = \int_0^1 dx_1\int_0^1dx_2\,\frac{15 x_2^2}{\left(x_1^2 + x_2^2 + 1\right)^{7/2}}. \end{equation}
The factor of $4$ in Eq.~\eqref{eq:dc} arises from exploiting evenness in $x_1$ and $x_2$ to restrict the double integral to the quadrant $[0,1]^2$.

Carrying out the integrations over $dx_2$ (using Eqs.~\eqref{eq:i5} to~\eqref{eq:i27}) yields
\begin{equation} J_5-K_2 = \int_0^1 dx_1\, \frac{3}{\left(2 + x_1^2 \right)^{5/2}} = \frac{2}{3\sqrt{3}} \label{eq:j5k2}, \end{equation}
and
\begin{equation}J_7-K_2 = \int_0^1 dx_1\, \frac{22+14x_1^2}{3(1+x_1^2)^3 (2+x_1^2)^{5/2} } = \frac{10}{9\sqrt{3}} \label{eq:j7k2}, \end{equation}
where the remaining integrals are evaluated analytically using Eqs.~\eqref{eq:i5} and~\eqref{eq:i35}.
Substituting Eqs.~\eqref{eq:j5k2} and~\eqref{eq:j7k2} into Eq.~\eqref{eq:dc} gives
\begin{equation} D_{xy} = \frac{24 r^2 z^2 - 40 z^4}{9\sqrt{3}}, \end{equation}
and cyclic permutation yields
\begin{equation} D_{yz} = \frac{24 r^2 x^2 - 40 x^4}{9\sqrt{3}};\quad  D_{zx} = \frac{24 r^2 y^2 - 40 y^4}{9\sqrt{3}} .\end{equation}
Substituting these results into Eq.~\eqref{eq:i0d} yields the expression stated in Eq.~\eqref{eq:inti0}, completing the derivation.
Finally, using $r^4 = x^4 + y^4 + z^4 + 2(x^2 y^2 + y^2 z^2 + z^2 x^2)$, the result may be recast as an alternative linear combination of quadrupolar invariants:
\begin{equation} I_0 = \frac{ 80\left( x^2 y^2 + y^2 z^2 + z^2 x^2\right) - 16r^4 }{9\sqrt{3}}. \end{equation}

We have explicitly evaluated the integral over the cubic domain and expressed the result as linear combinations of quadrupolar terms. 
For a general cuboid domain, the same structural form holds, with coefficients given by analogous double integrals; however, their explicit evaluation is more involved and lies beyond the scope of this work.
\end{document}